\documentclass[tightenlines,preprint,amssymb,amsmath,aps,prd,dvips,draft,showpacs]{revtex4}
\usepackage[final]{graphicx}

\newcommand{\calA}{\mathcal{A}}
\newcommand{\calB}{\mathcal{B}}
\newcommand{\calC}{\mathcal{C}}
\newcommand{\calE}{\mathcal{E}}
\newcommand{\calF}{\mathcal{F}}
\newcommand{\calH}{\mathcal{H}}
\newcommand{\calL}{\mathcal{L}}
\newcommand{\calP}{\mathcal{P}}
\newcommand{\calR}{\mathcal{R}}
\newcommand{\calX}{\mathcal{X}}
\newcommand{\calY}{\mathcal{Y}}
\newcommand{\calZ}{\mathcal{Z}}

\newcommand{\gotg}{\mathfrak{g}}
\newcommand{\new}{{\text{new}}}
\newcommand{\old}{{\text{old}}}

\newcommand{\tpmtpp}{{t_p-t_\pp}}
\newcommand{\lm}{{\ell m}}
\renewcommand{\l}{{\ell}}
\newcommand{\ro}{{r_{\text{o}}}}
\newcommand{\M}{M}

\newcommand{\Epsi}{{\epsilon}}

\newcommand{\XX}{\calX}
\newcommand{\YY}{\calY}
\newcommand{\ZZ}{\calZ}

\newcommand{\x}{x}
\newcommand{\y}{y}
\newcommand{\z}{z}
\renewcommand{\t}{t}

\newcommand{\xx}{\tilde{x}}
\newcommand{\yy}{\tilde{y}}
\newcommand{\zz}{\tilde{z}}

\newcommand{\Om}{\Omega}

\newcommand{\stf}{{\text{STF}}}
\newcommand{\tail}{{\text{tail}}}
\newcommand{\self}{{\text{self}}}
\newcommand{\dir}{{\text{dir}}}
\newcommand{\ret}{{\text{ret}}}
\newcommand{\adv}{{\text{adv}}}

\newcommand{\sym}{{\text{sym}}}
\newcommand{\R}{{\text{R}}}
\renewcommand{\SS}{{\text{S}}}
\newcommand{\rhoB}{\tilde{\rho}}

\newcommand{\HB}{\bar{H}}

\newcommand{\sgn}{{\text{sgn}}}
\newcommand{\p}{\prime}
\newcommand{\pp}{{p^\prime}}
\newcommand{\ppp}{p,p^\prime}
\newcommand{\s}{{\text{s}}}
\newcommand{\Lie}{\calL}

\newcommand{\Deqn}[1]{{Eq.~(\ref{#1})}}
\newcommand{\Deqns}[1]{{Eqs.~(\ref{#1})}}

\newcommand{\be}{\begin{equation}}
\newcommand{\ee}{\end{equation}}

\newcommand{\beq}{\begin{equation}}
\newcommand{\eeq}{\end{equation}}

\newcommand{\bea}{\begin{eqnarray}}
\newcommand{\eea}{\end{eqnarray}}

\begin{document}
\title{Self-force of a scalar field for circular
           orbits about a Schwarzschild black hole}
\author{Steven Detweiler}
\author{Eirini Messaritaki}
\author{Bernard F. Whiting}
\affiliation{Department of Physics, PO Box 118440, University of Florida,
          Gainesville, FL 32611-8440}
\date{February 20, 2003}

\begin{abstract}

The foundations are laid for the numerical computation of the actual
worldline for a particle orbiting a black hole and emitting gravitational
waves. The essential practicalities of this computation are here
illustrated for a scalar particle of infinitesimal size and small but
finite scalar charge. This particle deviates from a geodesic because it
interacts with its own retarded field $\psi^\ret$. A recently introduced
\cite{DetWhiting02} Green's function $G^\SS$ precisely determines the
singular part, $\psi^\SS$, of the retarded field. This part exerts no
force on the particle. The remainder of the field $\psi^\R = \psi^\ret -
\psi^\SS$ is a vacuum solution of the field equation and is entirely
responsible for the self-force. A particular, locally inertial coordinate
system is used to determine an expansion of $\psi^\SS$ in the vicinity of
the particle. For a particle in a circular orbit in the Schwarzschild
geometry, the mode-sum decomposition of the difference between $\psi^\ret$
and the dominant terms in the expansion of $\psi^\SS$ provide a mode-sum
decomposition of an approximation for $\psi^\R$ from which the self-force
is obtained. When more terms are included in the expansion, the
approximation for $\psi^\R$ is increasingly differentiable, and the
mode-sum for the self-force converges more rapidly.

\end{abstract}
 \maketitle

\section{Introduction}
 \label{Intro}

In general relativity, a particle of infinitesimal mass will orbit a black
hole of large mass along a worldline $\Gamma$ which is an exact geodesic
in the background geometry determined by the large mass alone.  If the
orbiting particle is not infinitesimal, having a small finite mass, its
orbit will no longer be a geodesic in the background of the larger mass,
and gravitational waves will be emitted by the system --- at infinity. In
the neighborhood of the small particle, local measurements cannot
separately distinguish the background of the large mass from a
\textit{smooth} perturbation to it caused by the presence of the smaller
mass \cite{DetWhiting02}. The actual orbit of the particle can be analyzed
in a linearization of the Einstein equations via a perturbation expansion
in the ratio of the masses. Through first order in this ratio, $\Gamma$ is
known to be a geodesic of a geometry perturbed from the background of the
large mass by the presence of the smaller one \cite{DetWhiting02}. The
difference of the worldline from a geodesic in the background is said to
arise from the interaction of the orbiting particle with its own
gravitational field.  It is said to result from a ``self-force,'' even
though, in the perturbed geometry determined by both the small and large
masses, the orbit would be observed to be geodesic.

In a strict sense, if the particle is of infinitesimal size, then its own
field is singular along its worldline, and there the perturbation analysis
fails. This difficulty can be avoided by allowing the size of the particle
to remain finite while invoking the conservation of the stress-energy
tensor within a world-tube which surrounds the worldline, in a manner
similar to Dirac's \cite{Dirac38} classical analysis. The balance of
energy and momentum indicates how to calculate the self-force in a way
which is independent of the size of the particle. The limit of vanishing
size may then be taken without confusion.

In curved spacetime, analyses beginning with DeWitt and Brehme
\cite{DeWittBrehme60} and subsequently by Mino, Sasaki and Tanaka
\cite{Mino97} and by Quinn and Wald \cite{QuinnWald97,Quinn00} formally
resolve the difficulty presented by the singularity in curved spacetime
with a Hadamard expansion \cite{DeWittBrehme60} of the Green's function
near $\Gamma$. The retarded Green's function $G^\ret(p,\pp)$ incorporates
physically appropriate boundary conditions and describes the field
$\psi^\ret$ of a particle moving through a given spacetime. In most past
discussions of the self-force, $G^\ret(p,\pp)$ is commonly divided into a
``direct'' part, which has support only on the past null cone of the field
point $p$, and a ``tail'' part, which has support inside the past null
cone and is a result of the curvature of spacetime. The analyses show that
$\psi^\tail = \psi^\ret - \psi^\dir$ is necessarily finite at the particle
and suggest that it is the part of $\psi$ which belongs on the right hand
side of an equation for the self-force
 \cite{Quinn00}
\beq
   \calF_a = q \nabla_a \psi .
\label{scalarforce}
\eeq
In these approaches, the use of $\psi^\tail$ instead of $\psi^\ret$
constitutes a form of regularization of the singular $\psi^\ret$.
Actually, while finite, $\psi^\tail$ is generally not differentiable on
the worldline \cite{DeWittBrehme60} if the Ricci scalar of the background
is not zero. Similarly, the electromagnetic potential $A^\tail_a$
(respectively, the gravitational metric perturbation $h^\tail_{ab}$) is
not differentiable at the particle if $(R_{ab}-\frac{1}{6}g_{ab}R) u^b$
(respectively, $R_{cadb}u^c u^d$) is nonzero in the background. In all
such cases, some version of averaging must be invoked to make sense of the
self-force. Moreover, we find it instructive to observe that the tail part
of the field is necessarily associated with a nonphysical inhomogeneous
source, i.e. $\nabla_a\nabla^a \psi^\tail \ne 0$: cf. \Deqn{DelPsi}.

In this paper an alternative regularization of the field $\psi^\ret$ is
used to compute the self-force where, in particular, by regularization we
mean not only controlling the singular behavior, but also the
differentiability. We have recently given a precise procedure for
decomposing the retarded field in neighborhood of $\Gamma$ into two parts
\cite{DetWhiting02}
\beq
  \psi^\ret = \psi^\SS + \psi^\R,
\label{defpsiR}
\eeq
where $\psi^\SS$ is a solution of the inhomogeneous field equation for the
particle, and is determined in the neighborhood of the worldline entirely
by local analysis via \Deqn{greenSS}. As both $\psi^\ret$ and $\psi^\SS$
are inhomogeneous solutions of the same differential equation, it follows
that $\psi^\R$, defined by \Deqn{defpsiR} is necessarily a homogeneous
solution and is therefore expected to be differentiable on $\Gamma$. In
Ref.~\cite{DetWhiting02} we showed that $\psi^\R$ formally gives the
correct self-force when substituted on the right hand side of
\Deqn{scalarforce} in place of $\psi^\tail$. In this paper $\psi^\R$ is
used for an explicit computation of the self-force. We consider $\psi^\SS$
to be associated with the \textbf{S}ingular \textbf{S}ource, and $\psi^\R$
with the \textbf{R}egular \textbf{R}emainder.

While the procedure which follows from \Deqn{defpsiR} is well understood
in principle, its application to physically interesting situations remains
a challenge. In this paper we consider a particle endowed with a scalar
charge $q$ in circular motion about a Schwarzschild black hole.  On a
technical level, a spherical harmonic decomposition of both $\psi^\ret$
and $\psi^\SS$ provides the multipole components of each, and the mode by
mode sum of the difference of these components determines $\psi^\R$ and,
thence, the self-force.

In Section II we give a brief overview of the relation between our work
and that of earlier authors. We also summarize our analytical results and
introduce the additional regularizing parameters which allow us to obtain
increased convergence in our mode sum representation of the self-force.

A special set of coordinates is described in Section \ref{THZcoords}; the
THZ coordinates, introduced by Thorne and Hartle \cite{ThorneHartle85} and
extended by Zhang \cite{Zhang86}, are locally inertial on a geodesic.
These coordinates are convenient for describing the scalar wave equation
in the vicinity of the geodesic, where the metric takes a particularly
advantageous form. In Section \ref{GreensFunction} the Hadamard expansion
for the Green's function, discussed in detail by DeWitt and Brehme
\cite{DeWittBrehme60}, is described in terms of Synge's \cite{Synge60}
``world function'' $\sigma(p,\pp)$, which is defined as half of the square
of the geodesic distance between two points $p$ and $\pp$. We obtain both
$\sigma(p,\pp)$ and the Hadamard expansion in terms of the THZ
coordinates.

Section \ref{RegularizationP} outlines the determination of the
regularization parameters given below in \Deqns{Firstterm} to
(\ref{Lastterm}).  These results are in agreement with, but extend by
going to higher order, the work of Barack and Ori
\cite{BarackOri00,BarackOri02,BMNOS02} and Mino, Nakano and Sasaki
\cite{BMNOS02,Mino02}.

In Section \ref{Fitting}, with a concrete application of our method, we
examine a scalar charge in a circular orbit of the Schwarzschild geometry
at a radius of $10M$. It is in this section that we see the practical
advantage of using a higher order approximation in the regularization of
$\psi^\SS$. The additional parameters we find enable us to increase
dramatically the rate of convergence in the self-force summation.

In several Appendices we include details concerning the THZ coordinates,
the mathematical analyses which focus on calculation of the regularization
parameters and a brief summary of details concerning the integration of
the scalar wave equation in the Schwarzschild geometry.

\subparagraph{Notation:} $a,b,\ldots$ are four dimensional space-time
indices. $i,j,\ldots$ are three dimensional spatial indices.
$(t_\s,r,\theta,\phi)$ are the usual Schwarzschild coordinates. $t,x,y,z$
are locally-inertial THZ coordinates attached to the geodesic $\Gamma$,
and $\rho^2\equiv x^2+y^2+z^2$. The geodesic $\Gamma$ is given as
$x^a=z^a(\tau)$, where $\tau$ is the proper time along $\Gamma$. The flat
spatial metric in Cartesian coordinates is $\delta_{ij}$. The flat
Minkowski metric in Minkowski coordinates is $\eta_{ab}=(-1,1,1,1)$. The
points $p$ and $\pp$ refer to a field point and a source point on the
world line of the particle, respectively. In the coincidence limit
$p\rightarrow \pp$. An expression such as $O(\rho^n)$ means of the order
of $\rho^n$ as $x^i\rightarrow0$ in the THZ coordinates. But note that the
differentiability of such an order term is only necessarily $C^{n-1}$ at
$x^i=0$.

\section{Overview of relation to earlier work}
\label{Overview}

Formally, although our approach differs from that of Barack and Ori, our
method of implementation is similar to that in their pioneering analysis
in Ref.~\cite{BarackOri00,BarackOri02,BarackOri03}.  In their procedure,
which Burko has implemented \cite{Burko00,BarackBurko00} both for a scalar
field with radial and with circular orbits of Schwarzschild, the
self-force may be thought of as being evaluated from \footnote{We use
$\psi^\ret$ throughout in places where other authors have used
$\psi^{\text{full}}$ or $\psi^{\text{total}}$ to denote the ``actual''
field \cite{Dirac38}.}
\beq
  \calF^\self_a = \lim_{p\rightarrow \pp}
  \left[ \calF_a^\ret(p) -\calF_a^\dir(p) \right] ,
\eeq
where $\pp$ is the event on $\Gamma$ where the self-force is to be
determined, $p$ is an event in the neighborhood of $\pp$, and the
relationship between $\calF_a(p)$ and $\psi(p)$ is as given in
\Deqn{scalarforce}. To make use of this equation, both $\calF_a^\ret(p)$
and $\calF_a^\dir(p)$ are expanded  into multipole $\l$-modes, with
$\calF_{\l a}^\ret(p)$ determined numerically. Typically the source is
expanded in terms of spherical harmonics, and then a similar expansion for
$\psi^\ret$ is used
\bea
  \psi^\ret &=&  \sum_{\lm} \psi^\ret_{\lm}(r,t) Y_\lm(\theta,\phi)
\label{psiretlm}
\eea
where $\psi^\ret_{\lm}(r,t)$ is found numerically.  The individual $\lm$
components of $\psi^\ret$ in this expansion are finite at the location of
the particle even though their sum is singular.  Then $\calF^\ret_{\l a}$
is finite and results from summing $q\nabla_a(\psi^\ret_{\lm} Y_\lm)$ over
$m$. The $\l$-mode expansion of $\calF_{a}^\dir(p)$ was initially
determined by a local analysis of the Green's function for an orbit at
$\ro$ in Schwarzschild coordinates in Ref.~\cite{BarackOri00},
\beq
 \lim_{r\rightarrow\ro} \calF^\dir_{\l a}
    =   \left( \l+\frac12 \right)A_a + B_a
          + \frac{C_a}{\l+\frac12} + O(\l^{-2}),
\label{Fabco}
\eeq
in which it was found that the $O(\l^{-2})$ terms yield precisely zero
when summed over $\l$. Moreover, for circular geodesics in the equatorial
plane of the Schwarzschild geometry, the regularization parameter $C_a=0$
and $A_a$ and $B_a$ also vanish except for their $r$ components. The
values of $A_r$ and $B_r$, first determined by Barack and Ori
\cite{BarackOri00,BMNOS02,BarackOri02}, are given below in
\Deqns{Firstterm} and (\ref{Secondterm}). A further term, which we shall
denote as $D_a^\p$, was also introduced in Ref.
\cite{BarackOri00,BMNOS02,BarackOri02} and shown there to be zero. It
refers to the sum of the $O(\l^{-2})$ terms in \Deqn{Fabco}.  We comment
further about the contribution of $D_a^\p$ towards the end of this
section. The self-force is ultimately calculated as
\beq
  \calF^\self_a =
  \sum_{\l=0}^\infty
    \left[ \lim_{p\rightarrow \pp} \calF_{\l a}^\ret
         -  \left(\l+\frac12\right)A_a - B_a - \frac{C_a}{\l+\frac12}
    \right] + D_a^\p .
\label{FselfABC}
\eeq
Burko \cite{Burko00} notes in his numerical analysis that the terms in
this sum scale as $1/\l^2$ for large $\l$, the sum converges as $1/\l$,
and it is evident from his results that he computes to at least $\l=80$
and finds improved convergence with Richardson extrapolation.

From our perspective, the self-force at a point $\pp$ on $\Gamma$ is
formally given by
\beq
  \calF^\self_a =  \calF^\ret_a - \calF^\SS_a
                =  \calF^\R_a  \equiv q\nabla_a\psi^\R
\label{Fselfret-S}
\eeq
evaluated at the source point $\pp$. Formally, the function $\psi^\SS$ is
defined only in a neighborhood of $\pp$; however for calculational
purposes, the function may be extended in any smooth manner throughout the
spacetime. While the spherical harmonic components of this extended
function in the Schwarzschild geometry are not uniquely determined, they
still provide a convergent expression for $\psi^\SS$ for events near
$\pp$. Thus, in the Schwarzschild geometry the spherical harmonic
expansions of $\psi^\SS$ and $\psi^\ret$ yield
\beq
 \psi^\R_\lm(r,t) = \psi^\ret_\lm(r,t)  - \psi^\SS_\lm(r,t),
\eeq
and the self-force can be determined by evaluating the vector field
\bea
   \calF_a^\self &=& \nabla_a\sum_\lm \psi^\R_\lm Y_\lm
\nonumber\\ &=&
  \nabla_a \sum_\lm (\psi^\ret_\lm  - \psi^\SS_\lm)Y_\lm
\eea
at the source point $\pp$. Further, with the definitions
\beq
  \calF^{\SS / \ret}_{\l a} = \nabla_a \sum_m \psi^{\SS / \ret}_\lm Y_\lm ,
\eeq
the self-force is
\bea
   \calF_a^\self &=& \sum_\l \left(  \calF^\ret_{\l a} -
         \calF^\SS_{\l a}\right)
\label{sumoverl}
\eea
evaluated at $\ro$. In the above expressions the difference in multipole
moments must be taken before the summation over $\l$.

In our approach the regularization parameters are derived from the
multipole components of $\nabla_a\psi^\SS$ evaluated at the source point
and are used to control both singular behavior and differentiability. In
Section \ref{RegularizationP} we consider circular orbits of the
Schwarzschild geometry at radius $\ro$ and show that
\bea
 \lim_{r\rightarrow \ro} \calF^\SS_{\l r}
    &=&
       \left(\l+\frac12\right)A_r + B_r
       -  \frac{2\sqrt{2}D_r}{(2\l-1)(2\l+3)}
\nonumber\\ && \qquad
       {} +  \frac{E^1_r \calP_{3/2}}{(2\l-3)(2\l-1)(2\l+3)(2\l+5)} + O(\l^{-6}) .
\label{FABDE}
\eea
where the regularization parameters are independent of $\l$ and given by
\beq
  A_r  =
    - \sgn(\Delta) \frac{[\ro(\ro-3M)]^{1/2}}{\ro^2(\ro-2M)}
\label{Firstterm}
\eeq
\beq
 B_r = - \left[\frac{\ro-3M}{\ro^4(\ro-2M)}\right]^{1/2}
           \left[F_{1/2} - \frac{(\ro-3M)F_{3/2}}{2(\ro-2M)} \right],
\label{Secondterm}
\eeq
and
\bea
 D_r &=& \left[ \frac{2\ro^2(\ro-2M)}{\ro-3M}\right]^{1/2}
  \left[ -\frac{M(\ro-2M)F_{-1/2}}{2\ro^4(\ro-3M)}
       - \frac{(\ro-M)(\ro-4M)F_{1/2}}{8\ro^4(\ro-2M)} \right.
\nonumber\\ && \left.
      {} + \frac{(\ro-3M)(5\ro^2 - 7\ro M - 14M^2)F_{3/2} }{16\ro^4(\ro-2M)^2}
       - \frac{3(\ro-3M)^2(\ro+M)F_{5/2}}{16\ro^4(\ro-2M)^2} \right] .
\label{Lastterm}
\eea
$E^1_r$ has not yet been determined analytically, but the constant
$\calP_{3/2}$ in \Deqn{FABDE} is independent of $\l$ and is given in
\Deqn{Bk+1/2}; the $\l$ dependence of the $E^1_r$ term, and of higher
order parameters ($E^k_r$, $k>1$), is discussed in Section
\ref{RegularizationP}. In these expressions $F_q$ refers to the
hypergeometric function ${}_2F_1[q,\frac12;1;M/(\ro-2M)]$. The $A_r$ and
$B_r$ terms agree with the results of references
\cite{BarackOri00,BMNOS02,Mino02,BarackOri02} restricted to circular
orbits. When summed over all $\l$, the $D_r$ and $E^k_r$ terms
individually give no contribution to the self-force. This is consistent
with the results in \cite{BarackOri00,BarackOri02,BMNOS02,Mino02}, but
note the different definition of $D_r$ there, which we have referred to
above as $D_a^\p$. Our results thus yield the identical self-force to that
of Barack and Ori.

As we shall show in Section IV, in general $\psi^\SS$ can be known only
approximately. If we ignore the $D_r$ and $E^k_r$ terms in the
approximation for $\psi^\SS$, then the approximation for $\psi^\R$ is only
$C^1$. Hence, the $D_r$ and $E^k_r$ terms must be included for $\psi^\R$
to be a homogeneous solution of \Deqn{DelPsi} as discussed above. Although
we have just indicated above that the $D_r$ and $E^k_r$ terms give no
overall contribution to the self-force, we find that understanding the
nature of these additional terms can be used to speed up dramatically the
convergence of the sum in \Deqn{sumoverl}. We have used this understanding
in obtaining the results of Section \ref{Numerical}.

\section{THZ Normal coordinates}
 \label{THZcoords}

The scalar wave equation takes a simple form when written in a particular
coordinate system in which the background geometry looks as flat as
possible. Consider a geodesic $\Gamma$ through a background vacuum
spacetime geometry $g_{ab}$. Let $\calR$ be a representative length scale
of the background geometry---the smallest of the radius of curvature, the
scale of inhomogeneities, and the time scale for changes in curvature
along $\Gamma$. A {\it normal} coordinate system can always be found
\cite{MTW} where, on $\Gamma$, the metric and its first derivatives match
the Minkowski metric, and the coordinate $t$ measures the proper time.
Normal coordinates for a geodesic are not unique, and we use particular
coordinates which were introduced by Thorne and Hartle
\cite{ThorneHartle85} and extended by Zhang \cite{Zhang86} to describe the
external multipole moments of a vacuum solution of the Einstein equations.
In Appendix \ref{FindTHZcoords}, we give a constructive algorithm for
finding these THZ coordinates for any particular geodesic in a vacuum
spacetime. In THZ coordinates
\begin{eqnarray}
  g_{ab} &=& \eta_{ab} + H_{ab}
\nonumber\\
     &=&  \eta_{ab} + {}_2H_{ab} + {}_3H_{ab} + O(\rho^4/\calR^4),
                \; \rho/{\cal R}\rightarrow0,
\label{H's}
\end{eqnarray}
with
\begin{eqnarray}
  {}_2H_{ab}d\x^a d\x^b & = &
         - \calE_{ij} \x^i \x^j ( d\t^2 + \delta_{kl} d\x^k d\x^l )
\nonumber\\ &&
         {} + \frac{4}{3} \epsilon_{kpq}\calB^q{}_i \x^p \x^i d\t\, d\x^k
\nonumber\\ &&
         {} - \frac{20}{21} \left[ \dot{\calE}_{ij}x^ix^j x_k
            - \frac{2}{5} \rho^2 \dot{\calE}_{ik} x^i \right] d\t\, d\x^k.
\nonumber\\ &&
 {} + \frac{5}{21} \left[ x_i \epsilon_{jpq} \dot{\calB}^q{}_k x^px^k
    -  \frac{1}{5} \rho^2 \epsilon_{pqi}
                 \dot{\calB}_{j}{}^q x^p \right] \,d\x^i\, d\x^j.
\label{H2}
\end{eqnarray}
and
\begin{eqnarray}
  {}_3H_{ab}d\x^a d\x^b & = &
         - \frac13\calE_{ijk} \x^i \x^j \x^k ( d\t^2 + \delta_{kl} d\x^k d\x^l )
\nonumber\\ &&
         {} + \frac{2}{3} \epsilon_{kpq}\calB^q{}_{ij} \x^p \x^i\x^j d\t\, d\x^k
          +  O(\rho^4/\calR^4)_{ij} \, dx^i \, dx^j,
\label{H3}
\end{eqnarray}
where $\eta_{ab}$ is the flat Minkowski metric in the THZ coordinates
$(t,x,y,z)$, $\epsilon_{ijk}$ is the flat space Levi-Civita tensor,
$\rho^2=x^2+y^2+z^2$ and the indices $i$, $j$, $k$, $l$, $p$ and $q$ are
spatial and raised and lowered with the three dimensional flat space
metric $\delta_{ij}$. Note that a term of $O(\rho^4/\calR^4)$ is only
known to be $C^3$ in the limit. We call coordinates where $H_{ab}$ matches
only \Deqn{H2} second order THZ coordinates; these coordinates are well
defined up to the addition of arbitrary functions of $O(\rho^4/\calR^3)$.
Third order THZ coordinates match \Deqn{H's} through the terms in
\Deqn{H3}; these are well defined up to the addition of arbitrary
functions of $O(\rho^5/\calR^4)$. The external multipole moments are
spatial, symmetric, tracefree tensors and are related to the Riemann
tensor evaluated on $\Gamma$ by
\beq
 \calE_{ij} = R_{titj},
\eeq
\beq
  \calB_{ij} = \epsilon_i{}^{pq}R_{pqjt}/2,
\eeq
\beq
  \calE_{ijk} =\left[\nabla_kR_{titj}\right]^\stf
\eeq and
\beq
   \calB_{ijk} = \frac38 \left[\epsilon_i{}^{pq}\nabla_kR_{pqjt}\right]^\stf .
\eeq
where \stf\ means to take the symmetric, tracefree part with respect to
the spatial indices $i$, $j$ and $k$. $\calE_{ij}$ and $\calB_{ij}$ are
$O(1/\calR^2)$, and $\calE_{ijk}$ and $\calB_{ijk}$ are $O(1/\calR^3)$.
The dot denotes differentiation of the multipole moment with respect to
$t$ along $\Gamma$. That all of the above external multipole moments are
tracefree follows from the assumption that the background geometry is a
vacuum solution of the Einstein equations.

The THZ coordinates are a specialization of harmonic coordinates, and it
is useful to define the ``Gothic'' form of the metric
\beq
  \gotg^{ab} \equiv \sqrt{-g} g^{ab}
\label{gotgdef}
\eeq
as well as
\beq
  \HB^{ab} \equiv \eta^{ab} - \gotg^{ab} .
\label{HBdef}
\eeq
A coordinate system is harmonic if and only if
\beq
  \partial_a \HB^{ab} = 0.
\label{HBdiv}
\eeq
Zhang \cite{Zhang86} gives an expansion of $\gotg^{ab}$ for an arbitrary
solution of the vacuum Einstein equations in THZ coordinates, his
equation~(3.26). The lower order terms of $\HB^{ab}$ in this expansion are
\beq
 \HB^{ab} = {}_2\HB^{ab} + {}_3\HB^{ab} + O(\rho^4/\calR^4)
\eeq
where
\begin{eqnarray}
  {}_2\HB^{tt}& = & - 2 \calE_{ij} \x^i \x^j
\nonumber\\
  {}_2\HB^{tk}& = & - \frac{2}{3} \epsilon^{kpq}\calB_{qi}\x_p \x^i
         {} + \frac{10}{21} \left[ \dot{\calE}_{ij}x^ix^j x^k
            - \frac{2}{5} \dot{\calE}_{i}{}^k x^i \rho^2\right]
\nonumber\\
  {}_2\HB^{ij}& = & \frac{5}{21} \left[ x^{(i}
       \epsilon^{j)pq} \dot{\calB}_{qk} x_p x^k
    -  \frac{1}{5} \epsilon^{pq(i} \dot{\calB}^{j)}{}_q x_p \rho^2 \right]
\label{HB2}
\end{eqnarray}
and
\begin{eqnarray}
  {}_3\HB^{tt} & = &
         - \frac23\calE_{ijk} \x^i \x^j \x^k
\nonumber\\
  {}_3\HB^{tk} & = & - \frac{1}{3} \epsilon^{kpq}\calB_{qij} \x_p \x^i\x^j
\nonumber\\
  {}_3\HB^{ij} & = & O(\rho^4/\calR^4) .
\label{HB3}
\end{eqnarray}

At linear order in $\HB^{ab}$, the metric perturbation $H_{ab}$ is the
trace reversed version of $\HB^{ab}$,
\beq
H_{ab} = \HB_{ab} - \frac12 g_{ab} \HB^c{}_c ,
\eeq
and \Deqns{H's}-(\ref{H3}) are precisely the terms up to
$O(\rho^4/\calR^4)$ which correspond to Zhang's \cite{Zhang86} expansion.

\section{Green's functions for a scalar field}
 \label{GreensFunction}

The scalar field equation
\begin{equation}
   \nabla^2 \psi = -4\pi\varrho
\label{DelPsi}
\end{equation}
is formally solved in terms of a Green's function,
\begin{equation}
  \nabla^2 G(\ppp) = - (-g)^{-1/2} \delta^4(x^a_p-x^a_\pp) ,
\label{Greenseqn}
\end{equation}
where $\pp$ represents a source point on $\Gamma$, and $p$ a nearby field
point.  The source function for a point charge moving along a worldline
$\Gamma$, described by $\pp(\tau)$, is
\begin{eqnarray}
   \varrho(p) &=& q \int (-g)^{-1/2} \delta^4(p-\pp(\tau)) \,d\tau ,
\label{rhox}
\end{eqnarray}
where $\tau$ is the proper time along the worldline of the particle with
scalar charge $q$. The scalar field of this particle is
\begin{equation}
  \psi(p) = 4\pi q \int G[p,\pp(\tau)] \,d\tau .
\end{equation}

DeWitt and Brehme \cite{DeWittBrehme60} analyze scalar-field self-force
effects by using the Hadamard expansion of the Green's function. An
important quantity is Synge's \cite{Synge60} ``world function''
$\sigma(\ppp)$ which is half of the square of the distance along a
geodesic between two nearby points $p$ and $\pp$. The usual symmetric
scalar field Green's function is derived from the Hadamard form to be
\cite{DeWittBrehme60}
\beq
  G^\sym(\ppp) = \frac{1}{8\pi} \left[ u(\ppp) \delta(\sigma)
         - v(\ppp) \Theta(-\sigma) \right]
\label{greensym}
\eeq
where $u(\ppp)$ and $v(\ppp)$ are bi-scalars described by DeWitt and Brehme.
The $\Theta(-\sigma)$ guarantees that only when $p$ and $p^\p$ are timelike
related is there a contribution from $v(\ppp)$.

The retarded and advanced Green's functions are
\bea
  G^\ret(\ppp) &=& 2\Theta[\Sigma(p),\pp] G^\sym(\ppp)
\nonumber\\
  G^\adv(\ppp) &=& 2\Theta[\pp,\Sigma(p)] G^\sym(\ppp)
\label{greenRetAdv}
\eea
where $\Theta[\Sigma(p),p^\p]= 1 - \Theta[\pp,\Sigma(p)]$ equals 1 if
$p^\p$ is in the past of a spacelike hypersurface $\Sigma(p)$ that
intersects $p$, and is zero otherwise. The terms in a Green's function
containing $u$ and $v$ are commonly referred to as the ``direct'' and
``tail'' parts, respectively.

A second symmetric Green's function \cite{DetWhiting02}
\beq
  G^\SS(\ppp) = \frac{1}{8\pi} \left[ u(\ppp) \delta(\sigma)
         + v(\ppp) \Theta(\sigma) \right]
\label{greenSS}
\eeq
precisely identifies the part of $\psi$ which is not responsible for the
self-force.  In particular, the regular remainder
\beq
  \psi^\R = \psi^\ret - \psi^\SS
\eeq
is a homogeneous solution of the the field equation (\ref{DelPsi}) and
completely provides the self-force when put on the right hand side of
\Deqn{scalarforce} \cite{DetWhiting02}.

\subsection{Approximation for $\psi^\SS$}

In this section approximate expansions are derived for $G^\SS$ and
$\psi^\SS$. For a vacuum spacetime ($R_{ab}=0$) which is nearly flat
Thorne and Kov\'{a}cs \cite{ThorneKovacs75} show that
\beq
  u(\ppp) = 1 + O(\rho^4/\calR^4),
\label{udef}
\eeq
their equations~(39) and (40), and they evaluate the direct part of the
retarded Green's function to be
\bea
 \frac{1}{4\pi} u(\ppp) \;\delta_\ret[\sigma(\ppp)]
    &=&
     \left(\frac{1 + O( \rho^4/\calR^4)}{4\pi
            \dot{\sigma}}\right)_{\tau_\ret} \delta(t_p-t_\ret),
\label{1green}
\eea
where the dot denotes a derivative with respect to $t_\pp$

We now express  $\dot{\sigma}_{\text{ret}}$ in terms of the THZ
coordinates to obtain \Deqn{dsigmadt} below.  When the source point $\pp$
is on $\Gamma$, $\sigma(\ppp)$ is particularly easy to evaluate in THZ
coordinates for $p$ close to $\pp$. Synge's \cite{Synge60} ``world
function'' $\sigma(\ppp)$ is shown by Thorne and Kov\'{a}cs
\cite{ThorneKovacs75} to be
\beq
  \sigma(\ppp) = \frac{1}{2} x^a x^b
        \left( \eta_{ab} + \int_{\calC} H_{ab}\;d\lambda \right)
        + O(\rho^6/\calR^4),
\label{1sigma}
\eeq
their equations~(37) and (38), where the THZ coordinates of $\pp$ are
$(t_\pp,0,0,0)$,  $\x^a$ is the coordinate of the field point $p$, and the
coordinates of the path of integration $\calC$ are given by
$\zeta^a(\lambda) = \lambda (\x^a-t_\pp\delta^a_t)$ with $\lambda$ running
from 0 to 1. We closely follow the analysis in \cite{ThorneKovacs75},
while using THZ coordinates, and only work through lower orders in
$\rho/\calR$.

Given $H_{ab}={}_2H_{ab}+{}_3H_{ab}$ from \Deqns{H2} and (\ref{H3}), the
integral of a component of $H_{ab}$ along $\calC$ is straightforward. For
example,
\bea
\int_{\calC} H_{tt} \;d\lambda &=&
     - \int_{\calC} \left(\calE_{ij} \zeta^i \zeta^j
     + \frac13\calE_{ijk} \zeta^i \zeta^j \zeta^k\right) \;d\lambda
     + O(\rho^4/\calR^4)
\nonumber\\ &=&
    - \frac13\calE_{ij} x^i x^j
    - \frac{1}{12}\calE_{ijk}  x^i x^j x^k + O(\rho^4/\calR^4) ,
\label{Eintegral}
\eea
The other components give similar results.  If we define
\beq
  \calH_{ab}  \equiv \int_{\calC} H_{ab} \;d\lambda ,
\label{calH}
\eeq
then Synge's world function is
\bea
 \sigma(\ppp) &=&
     \frac{1}{2}  x^a  x^b \eta_{ab}
      +  \frac12 (\tpmtpp)^2 \calH_{tt}
      +  (\tpmtpp) x^i \calH_{it}
      + \frac12  x^i x^j\calH_{ij}  + O(\rho^6/\calR^4),
\nonumber \\    &=& -\frac{1}{2}(1-\calH_{tt})
      \left[ \left(t_\pp - t_p +  x^i\calH_{it} \right)^2 -
              x^i  x^j (\eta_{ij} + \calH_{ij})/(1-\calH_{tt})
      \right]
\nonumber\\ &&
   \qquad \qquad {} + O(\rho^6/\calR^4).
\label{sigmaexp}
\eea
The second equality depends upon the facts that $\calH_{ab} =
O(\rho^2/\calR^2)$ and that $|t_\pp - t_p| = O(\rho)$ near the null cone.

With the source point on $\Gamma$,
\beq
   x^i  x^j (\eta_{ij}+\calH_{ij})
         = \rho^2(1+\calH_{tt}) + O(\rho^6/\calR^4)
\label{xxetaH}
\eeq
where
\beq
  \rho^2 =  x^i  x^j \eta_{ij}.
\eeq
The result in \Deqn{xxetaH} depends upon the detailed nature of
${}_2H_{ij}$ and ${}_3H_{ij}$ in \Deqns{H2} and (\ref{H3}) as well as upon
the definition of $\calH_{ij}$ in \Deqn{calH}.

After the substitution of \Deqn{xxetaH} into \Deqn{sigmaexp},
factorization of $\sigma$ yields
\bea
  \sigma(\ppp) &=& -\frac{1}{2}(1-\calH_{tt})
                     [t_{p^\p} - t_p +  x^i \calH_{it} - \rho(1+\calH_{tt})]
\nonumber\\ &&
     \quad {} \times  [t_{p^\p} - t_p +  x^i \calH_{it} + \rho(1+\calH_{tt})]
          + O(\rho^6/\calR^4) .
\label{sigmafact}
\eea
At the retarded time, $p^\p$ is on the past null cone emanating from
$p$, where $\sigma(\ppp)=0$, and it follows that the first of the
factors in square brackets is $\sim\rho$ and the second must be
$\rho^{-1}\times O(\rho^6/\calR^4)=O(\rho^5/\calR^4)$ to cancel the
order term in \Deqn{sigmafact} and have $\sigma(\ppp)$ vanish
precisely. Thus, differentiation of \Deqn{sigmafact} with respect to
$t_\pp$ and evaluation at the retarded time yields an expression
which is dominated by the part which results from the differentiation
of the second term in square brackets,
\bea
  \left[\frac{d\sigma(\ppp)}{dt_\pp}\right]_\ret  &=&
        -\frac{1}{2}(1-\calH_{tt})
                     [t_{p^\p} - t_p +  x^i \calH_{it} -
                     \rho(1+\calH_{tt})] + O(\rho^6/\calR^5)
\nonumber\\  &=&
        -\frac{1}{2}(1-\calH_{tt})
                     [- 2 \rho(1+\calH_{tt}) + O(\rho^5/\calR^4)]
\nonumber\\  &=&
         \rho [1 + O(\rho^4/\calR^4)];
\label{dsigmadt}
\eea
the first equality follows from taking the derivative of the second term
in square brackets in \Deqn{sigmafact} with respect to $t_\pp$, the second
equality from evaluating at the retarded time, and the third equality
follows from \Deqns{Eintegral} and (\ref{calH}).

The direct part of the retarded Green's function in \Deqn{1green} is
now
\begin{align}
 \frac{1}{4\pi} u(\ppp) \;\delta_\ret[\sigma(\ppp)]
   = &
     (4\pi \rho)^{-1}
       \delta[ t_{p^\p} - t_p  +  x^i \calH_{it}
\nonumber\\
     & + \rho(1+\calH_{tt}) + O(\rho^5/\calR^4)][1 + O(\rho^4/\calR^4)].
\label{dirGreen}
\end{align}

DeWitt and Brehme show that in general
\beq
  v(\ppp) = - \frac{1}{12} R(\pp) + O(\rho/\calR^3), \quad
       p \rightarrow \Gamma ,
\label{vcauchy}
\eeq
but in vacuum, where $R=0$ from the Einstein equations,
\beq
 v(\ppp) = O(\rho^2/\calR^4) .
\eeq
This follows from  Eq.~(2.9) with substitutions from
 Eqs.~(2.14), (2.15), (1.76) and (1.10) of Ref.~\cite{DeWittBrehme60}.
The dominant contribution to $\psi^\SS$ from the $v(\ppp)$ term is
$O(\rho^3/\calR^4)$ in the coincidence limit, $p\rightarrow \Gamma$.

All together then, with \Deqns{dirGreen} and (\ref{vcauchy}), substituted
into \Deqn{greenSS} and an integration over the worldline,
\beq
  \psi^\SS = q/\rho + O(\rho^3/\calR^4).
\label{psiSS}
\eeq
We note that the third order THZ coordinates are only well defined up to
the addition of a term of $O(\rho^5/\calR^4)$. Such an addition would
change $1/\rho$ by the sum of a term that is $O(\rho^3/\calR^4)$, and
would be consistent with the order term of \Deqn{psiSS}. The
differentiability of the order term is of interest, and a term of
$O(\rho^3/\calR^4)$ is $C^2$ in the limit that $\rho\rightarrow0$. In
light of the fact that $\psi^\R$ is a homogeneous solution, \Deqn{psiSS}
clarifies the relationship between the accuracy of an approximation for
$\psi^\SS$ and the differentiability of the subsequent approximation for
$\psi^\R = \psi^\ret - \psi^\SS$, and the self-force $\partial_a \psi^\R$.
Specifically, if the approximation for $\psi^\SS$ is in error by a $C^n$
function, then the approximation for $\psi^\R$ is no more differentiable
than $C^n$ and the approximation for $\partial_a \psi^\R$ is no more
differentiable than $C^{n-1}$.

\subsection{Intuitive understanding for $\psi^\SS$}

Before continuing, it is instructive to provide an elementary, direct
explanation of \Deqn{psiSS} by taking full advantage of the features
of the THZ coordinates. The scalar wave operator in THZ coordinates,
is
\beq
  \sqrt{-g}\nabla^a\nabla_a\psi =
         \partial_a \left( \eta^{ab} \partial_b \psi \right)
       - \partial_a \left( \HB^{ab}  \partial_b \psi \right)
\eeq
or
\beq
  \sqrt{-g}\nabla^a\nabla_a\psi =
         \eta^{ab} \partial_a \partial_b \psi
       -  \HB^{ij} \partial_i \partial_j \psi
       - 2 \HB^{it} \partial_{(i}  \partial_{t)} \psi
       -  \HB^{tt} \partial_t \partial_t \psi .
\label{del2psiHB}
\eeq
Direct substitution into \Deqn{del2psiHB} shows how well $q/\rho$
approximates $\psi^\SS$.  If $\psi$ is replaced by $q/\rho$ on the right
hand side, then the first term gives a $\delta$-function, the third and
fourth terms vanish because $\rho$ is independent of $t$, and in the second
term ${}_2\HB^{ij}$ has no contribution because of the details given in
\Deqn{HB2}, and the remainder of $\HB^{ij}$ yields a term that scales as
$O(\rho/\calR^4)$. Thus,
\beq
  \sqrt{-g}\nabla^a\nabla_a(q/\rho) = -4\pi q\delta^3(x^i)
          +  O(\rho/\calR^4), \quad \rho/\calR \rightarrow 0.
\label{del2rho}
\eeq
From consideration of solutions of Laplace's equation in flat
spacetime, it follows that a $C^2$ correction to $q/\rho$, of
$O(\rho^3/\calR^4)$, would remove the remainder on the right hand
side. We conclude that $\psi^\SS = q/\rho + O(\rho^3/\calR^4)$ is an
inhomogeneous solution of the scalar field wave equation. And the
error in the approximation of $\psi^\SS$ by $q/\rho$ is $C^2$.

\section{Regularization parameters for a circular orbit of the
         Schwarzschild geometry}
 \label{RegularizationP}

In Appendix \ref{ActualTHZ} we give the detailed functional relationship
between the THZ coordinates $(\t,x,y,z)$ and the Schwarzschild coordinates
$(t_\s,r,\theta,\phi)$ for an orbit described by $r=\ro$ with
$\theta=\pi/2$ and $\phi=\Omega t_\s$.

As seen in Section \ref{GreensFunction}, an approximation to $\psi^\SS$ is
\beq
  \psi^\SS = q/\rho + O(\rho^3/\calR^4) .
\eeq
The regularization parameters result from evaluating the multipole
components of $q/\rho$ at the location of the source.  The use in this
manner of $q/\rho$, in lieu of $\psi^\SS$ itself, is justified because the
error in the approximation to $\psi^\SS$, being $O(\rho^3/\calR^4)$, gives
no contribution to $\nabla_a\psi^\SS$ as $x\rightarrow0$.

To aid in the multipole expansion we rotate the usual Schwarzschild
coordinates to move the coordinate location of the particle from the
equatorial plane to a location where $\sin\Theta=0$ for a specific $t_\s$,
following the approach of Barack and Ori as described in
\cite{BarackOri02}.  Thus, we define new angles $\Theta$ and $\Phi$ in
terms of the usual Schwarzschild angles by
\bea
  \sin\theta\,\cos(\phi-\Omega t_\s) &=& \cos\Theta
\nonumber\\
  \sin\theta\,\sin(\phi-\Omega t_\s) &=& \sin\Theta\,\cos\Phi
\nonumber\\
  \cos\theta &=& \sin\Theta\,\sin\Phi .
\label{newangles}
\eea
A coordinate rotation maps each $Y_\lm(\theta,\phi)$ into a linear
combination of the $Y_{\l m^\p}(\Theta,\Phi)$ which preserves the index
$\l$, while $m^\p$ runs over $-\l\ldots\l$. Thus, the $\l$ component of the
self-force, after summation over $m$, is invariant under the coordinate
rotation.

To obtain the regularization parameters: first we expand $\partial_r
(q/\rho)$ into a sum of spherical harmonic components whose
amplitudes depend upon $r$. Then we take the limit $r\rightarrow\ro$.
Finally only the $m=0$ components contribute to the self-force at
$\Theta=0$ because $Y_\lm(0,\Phi)=0$ for $m\ne0$. Thus, the
regularization parameters of \Deqn{FABDE} are the $(\l,m=0)$
spherical harmonic components of $\partial_r(q/\rho)$ evaluated at
$\ro$.

In this section $\Epsi$ is a formal parameter which is to be set to unity
at the end of a calculation; a term containing a factor of $\Epsi^n$ is
$O(\rho^n)$.  We use it to help identify the behavior of certain terms in
the coincidence limit, $x^i\rightarrow0$. We have used \textsc{Maple} and
\textsc{grTensor} extensively to obtain the results reported below.

A lengthy expression for $\rho^2$, for a circular orbit in the
Schwarzschild geometry, may be derived from the analysis of Appendix
\ref{ActualTHZ}. The $O(\Epsi^2)$ part of $\rho^2$ is
\beq
 \rhoB^2 \equiv \frac{\ro\Delta^2}{\ro-2M}
          + 2 \ro^2 \frac{\ro-2M}{\ro-3M} \chi (1-\cos\Theta)
\label{rhoBdef}
\eeq
where
\beq
 \Delta\equiv(r-\ro),
\label{Deltadef}
\eeq
and
\beq
 \chi \equiv 1 - \frac{M\sin^2\Phi}{\ro - 2M}.
\label{chidef}
\eeq

The dependence of $\rho^2$ on the Schwarzschild coordinates may be written
solely in terms of $\Delta$, $\chi$ and $\rhoB$ by use of \Deqn{rhoBdef}
to remove $\cos\Theta$, then \Deqns{Deltadef} and (\ref{chidef}) remove
$r$ and $\sin\Phi$ respectively. We formally expand $\partial_r(1/\rho)$
in powers of $\Epsi$ to obtain
\bea
  \partial_r (1/\rho) &=& - \frac{\Epsi^{-2}\ro \Delta}{\rhoB^3(\ro-2M)}
      -  \frac{\Epsi^{-1}}{\rhoB \ro} \left[  1
             - \frac{\ro-3M}{2\chi(\ro-2M)}
                       + O(\Delta^2/\rhoB^2, \Delta^4/\rhoB^4) \right]
\nonumber\\ &&
   {} +\Epsi^0 O(\Delta/\rhoB,\Delta^2/\rhoB^2 )
      + \frac{\Epsi^1 \rhoB}{\ro^4} \left[ -\frac{M(\ro-2M)}{2(\ro-3M)}
       - \frac{(\ro-M)(\ro-4M)}{8\chi(\ro-2M)} \right.
\nonumber\\ && \left.
   {}  + \frac{(\ro-3M)(5\ro^2 - 7\ro M - 14M^2)}{16\chi^2(\ro-2M)^2}
    \right.
\nonumber\\ && \left.
   {}  - \frac{3(\ro-3M)^2(\ro+M)}{16\chi^3(\ro-2M)^2}
       + O(\Delta^2/\ro, \Delta^4/\rhoB^2\ro)
        \right]
       + O(\rho^2) .
\label{drhoinv}
\eea

We consider the multipole expansion, in the limit that
$\Delta\rightarrow0$, of the $m=0$ part of each coefficient of $\Epsi^n$
for $n=-2$ to $1$. A convenient method to find the $m=0$ component in one
of the following terms involves integrating over the angle $\Phi$ using
details described in Appendix \ref{MathEllip}. The expansion of the
$\Theta$ dependence in terms of Legendre polynomials is described in
Appendix \ref{MathLegendre}.

\subparagraph{First term} The $\Delta\rightarrow0$ limit of the
expansion of the $\Epsi^{-2}$ term in \Deqn{drhoinv} is
\bea
 \lefteqn{ \lim_{\Delta\rightarrow0}
          - \frac{\Epsi^{-2}\ro \Delta}{\rhoB^3(\ro-2M)}  } \quad & &
\nonumber \\  &=&
   \lim_{\Delta\rightarrow0} - \left(\frac{\ro}{\ro-2M}\right)^{1/2}
      \left(\frac{\ro \Delta^2}{\ro-2M} \right)^{1/2}
      \left[\frac{\ro\Delta^2}{\ro-2M} +
      \frac{2\ro^2(\ro-2M)}{\ro-3M}\chi(1-\cos\Theta) \right]^{-3/2}
\nonumber \\  &=&
    - \left(\frac{\ro}{\ro-2M}\right)^{1/2}
      \frac{\sgn(\Delta)(\ro-3M)}{\ro^2(\ro-2M-M\sin^2\Phi)}
      \sum_{\l=0}^\infty\left(\l+\frac12\right)P_\l(\cos\Theta),
\eea
where $\Epsi$ has been set equal to $1$ on the right hand side here and
below, and where the second equality follows from \Deqn{Al-3/2} with the
substitution
\beq
 \delta^2 = \Delta^2(\ro-3M)/[2\ro(\ro-2M)^2\chi].
\eeq
Integrating over $\Phi $ and dividing by $2\pi $ (denoted by the angle
brackets $\langle \rangle $ here and in Appendix~\ref{MathEllip})  via
\Deqn{F1}, to find the $m=0$ contribution, results in
\bea
 \lefteqn{ \lim_{\Delta\rightarrow0}
       \left\langle
             - \frac{\Epsi ^{-2}\ro \Delta}{\rhoB^3(\ro-2M)}
       \right\rangle }\quad & &
\nonumber \\  &=&
    - \sgn(\Delta) \frac{[\ro(\ro-3M)]^{1/2}}{\ro^2(\ro-2M)}
      \sum_{\l=0}^\infty\left( \l + \frac12 \right) P_\l(\cos\Theta) .
\label{firstterm}
\eea
In the coincidence limit $P_\l(\cos\Theta)=1$ and a term in this sum
is then
\beq
  - \left( \l + \frac12 \right) \sgn(\Delta)
      \frac{[\ro(\ro-3M)]^{1/2}}{\ro^2(\ro-2M)}
\eeq
which determines the $A_r$ term in \Deqn{FABDE} as given in
\Deqn{Firstterm}.

\subparagraph{Second term} The $\Delta\rightarrow0$ limit of the
$\Epsi^{-1}$ term in \Deqn{drhoinv} is
\bea
 \lefteqn{ \lim_{\Delta\rightarrow0} - \frac{\Epsi^{-1}}{\rhoB \ro}
     \left[1 - \frac{\ro-3M}{2\chi(\ro-2M)}  \right]}\quad &&
\nonumber\\
  &=&
    - \left[\frac{\ro-3M}{2\ro^4(\ro-2M-M\sin^2\Phi)(1-\cos\Theta)}
           \right]^{1/2}
           \left[1 - \frac{\ro-3M}{2\chi(\ro-2M)}  \right]
\nonumber\\
  &=&
    - \left[\frac{\ro-3M}{2\ro^4(\ro-2M)}\right]^{1/2} (1-\cos\Theta)^{-1/2}
           \left[\frac{1}{\chi^{1/2}} - \frac{\ro-3M}{2\chi^{3/2}(\ro-2M)}
           \right].
\eea
Integrating over $\Phi$ results in hypergeometric functions as shown in
\Deqn{s2ave}, and the expansion of $(1-\cos\Theta)^{-1/2}$ in terms of the
$P_\l(\cos\Theta)$ is given in \Deqn{Al-1/2} and results in
\bea
 \lefteqn{ \lim_{\Delta\rightarrow0} \left\langle - \frac{\Epsi^{-1}}{\rhoB \ro}
     \left[1 - \frac{\ro-3M}{\chi(\ro-2M)}  \right] \right\rangle }\quad &&
\nonumber\\
  &=&
    - \left[\frac{\ro-3M}{2\ro^4(\ro-2M)}\right]^{1/2}
           \left[F_{1/2} - \frac{(\ro-3M)F_{3/2}}{2(\ro-2M)}  \right]
      \sqrt{2}\sum_{\l=0}^\infty P_\l(\cos\Theta)
\label{secondterm}
\eea
In the coincidence limit $P_\l(\cos\Theta)=1$ and a term in this sum is
then
\beq
  B_r =  - \left[\frac{\ro-3M}{\ro^4(\ro-2M)}\right]^{1/2}
           \left[F_{1/2} - \frac{(\ro-3M)F_{3/2}}{\ro-2M}  \right]
\eeq
which is the $B_r$ term in \Deqn{FABDE} as given in \Deqn{Secondterm}.

\subparagraph{Third term} The $O(\Epsi^0)$ term in \Deqn{drhoinv} is zero in
the limit that $\Delta\rightarrow0$ for nonzero $\Theta$, and gives no
contribution to the sum in \Deqn{FABDE} as follows from \Deqn{Al-1/2}.

\subparagraph{Last term} For the last, $\Epsi^{1}$, term in \Deqn{drhoinv}
we consider
\beq
  \lim_{\Delta\rightarrow0} \rhoB
    = \left[ \frac{2\ro^2(\ro-2M)}{\ro-3M}\right]^{1/2} \chi^{1/2}
       (1-\cos\Theta)^{1/2} .
\eeq
After the expansion of $(1-\cos\Theta)^{1/2}$ described in \Deqn{Al+1/2}
and the integration over $\Phi$ with \Deqn{s2ave}, the multipole expansion
of the $\rhoB$ terms in \Deqn{drhoinv} gives
\bea &&
  \left[ \frac{2\ro^2(\ro-2M)}{\ro-3M}\right]^{1/2}
  \left[ -\frac{M(\ro-2M)F_{-1/2}}{2\ro^4(\ro-3M)}
       - \frac{(\ro-M)(\ro-4M)F_{1/2}}{8\ro^4(\ro-2M)} \right.
\nonumber\\ && \left.
       + \frac{(\ro-3M)(5\ro^2 - 7\ro M - 14M^2)F_{3/2} }{16\ro^4(\ro-2M)^2}
       - \frac{3(\ro-3M)^2(\ro+M)F_{5/2}}{16\ro^4(\ro-2M)^2} \right]
\nonumber\\ &&
  \times \sum_{\l=0}^\infty \frac{-2\sqrt{2}
         P_\l(\cos\Theta)}{(2\l-1)(2\l+3)}.
\label{lastterm}
\eea
In the coincidence limit $P_\l(\cos\Theta)=1$, and a term in this sum is then
\beq
   D_r \frac{-2\sqrt{2}}{(2\l-1)(2\l+3)},
\eeq
which defines $D_r$, as in \Deqn{Lastterm}, and is a term in \Deqn{FABDE}.

\subparagraph{The remainder} These derivations of the regularization
parameters reveal a pattern for the $\l$-dependence of higher order
parameters, even if the overall scale of the parameter remains unknown.

The successive terms in \Deqn{drhoinv} provide increasingly accurate
approximations of $\partial_r \psi^\SS$ and also increasingly accurate
approximations of $\partial_r \psi^\R=\partial_r \psi^\ret - \partial_r
\psi^\SS$. In principle the terms through $O(\Epsi^0)$ are sufficient to
calculate the radial component of the self-force; this is effectively the
level of approximation described in References
\cite{BarackOri00,BMNOS02,BarackOri02,Mino02} and implemented in
Ref.~\cite{Burko00}. With the inclusion of $O(\Epsi^0)$ terms the
approximation for $\partial_r\psi^\R$ is $C^0$, the remainder terms scale
as $\l^{-2}$ for large $\l$ and their sum converges. But when the
approximation of $\partial_r \psi^\SS$ is improved by the addition of the
$O(\Epsi)$ terms, the resulting approximation of $\partial_r \psi^\R$ is
then $C^1$, and we see below that the remainder terms scale as $\l^{-4}$
resulting in a more rapid convergence of the sum for the self-force.

As the approximation of $\partial_r \psi^\SS$ is improved by successive
terms of greater differentiability, the resulting approximation of
$\partial_r \psi^\R$ is not only more differentiable but also leads to
increasingly rapid convergence of the sum for the self-force.

We can anticipate the details of how this occurs.  From the descriptions
of the THZ coordinates in Appendix \ref{ActualTHZ} and of $\psi^\SS$ in
terms of THZ coordinates in Section \ref{GreensFunction}, we expect that
the $O(\Epsi^2)$ term in \Deqn{drhoinv} is $C^1$. A more accurate
approximation to $\psi^\SS$ could be provided by the modification
\beq
  \rho^2 \rightarrow \rho^2 + \lambda_N \calX^N,
\eeq
where the components of $\lambda_N = O(1/\calR^{n-2})$ are not functions
of the coordinates but depend only upon the orbit and the coordinate
location of the particle. Here we borrow the notation of the analysis of
STF tensors \cite{DamourIyer91} where $N$ is a multi-index that represents
$n$ spatial indices $i_1 i_2 \ldots i_n$, however while $\lambda_N$ is
symmetric it is not necessarily tracefree. Also  $\calX^N$ represents
$\calX^{i_1}\calX^{i_2} \ldots \calX^{i_n}$ where the $\calX^i$ represents
one of $\calX$, $\calY$ or $\calZ$ defined in Appendix \ref{ActualTHZ}.
Now, $\rho^2$ is already determined through $O(\Epsi^5)$, so that $n$ is
necessarily greater than or equal to 6 for an improved approximation. And
while we do not know the actual value of $\lambda_N$, we assume that such
a $\lambda_N$ exists that provides an improved approximation to
$\psi^\SS$.

Such a correction to $\rho^2$ ultimately results in the addition of
\beq
   \partial_r \left[ -\frac{\lambda_N\calX^N}{2\rhoB^3} + O(\rho^4/\calR^5) \right]
\label{partrcorrect}
\eeq
to $\partial_r(1/\rho)$ in \Deqn{drhoinv}, which is of $O(\Epsi^2)$ and
$\calC^1$ if $n=6$.  This result is consistent with \Deqn{psiSS} and with
the discussion following \Deqn{del2rho} in Section\ \ref{GreensFunction}
above. With this improved approximation the resulting error term in
\Deqn{drhoinv} would be $O(\Epsi^3)$ and $C^2$.

Finding higher order THZ coordinates might not be the best way to correct
the  approximation for $\psi^\SS$, but any such correction involving an
expansion about $\Gamma$ would necessarily take the generic form of a
homogeneous polynomial in $\calX$, $\calY$ and $\calZ$ (or equivalently in
$x^i$) divided by $\rhoB$ raised to some integral power. Thus to find
higher order corrections to $\psi^\SS$, we are led to consider the
multipole expansion of a term such as in \Deqn{partrcorrect}, for $n \ge
6$ and to determine the nature of the regularization parameters that would
result.

First $\calX$, $\calY$ and $\calZ$ are replaced by their definitions in
terms of the usual Schwarzschild coordinates in
\Deqns{calX}--(\ref{calZ}). Then, the angles are changed to $\Theta$ and
$\Phi$ via \Deqn{newangles}. And finally all coordinate dependence is
written in terms of $\Delta$, $\chi$ and $\rhoB$ as described above in
\Deqn{drhoinv}. The result is a sum of terms each of which is of
$O(\Epsi^{n-4})$, for $n \ge 6$ and whose coordinate dependence is
contained in the functions $\Delta$ (which depends only upon $r$), $\chi$
(which depends only upon $\Phi$) and $\rhoB$. As $x\rightarrow0$ both
$\Delta$ and $\rhoB$ are $O(\Epsi)$ while $\chi=O(1)$, so that each term
must include a factor $\Delta^q \rhoB^p$ for integers $q$ and $p$ where
$q+p=n-4 \ge 2$. In fact, careful analysis of the above substitutions
shows that $q \ge 0$. All of the $\Theta$ dependence resides in $\rhoB^p$.

The Legendre polynomial expansion of $\rhoB^p$, for odd $p$, is discussed
at length in Appendix \ref{MathLegendre}. There we show that when $p = -1$
(respectively, $p < -1$), a consequence of \Deqn{Al-1/2} [respectively,
\Deqn{A-k-1/2}] is that the expansion coefficients scale as a constant
(respectively, diverge as $(r-\ro)^p$) when $r \rightarrow \ro$. The
coefficients of the expansion $\Delta^q \rhoB^p$ then scale as
$(r-\ro)^{q}$ (respectively, $(r-\ro)^{q+p}$).  In both cases the power is
an integer $\ge 2$. In the limit that $r \rightarrow \ro$ these terms
approach zero and give no contribution to the regularization parameters.
The case of $p$ even and negative gives a similar result, but is not
discussed in the Appendix. We conclude that if $p<0$ then no contribution
to the regularization parameters results.

If $p \ge 0$ and $q>0$ then the Legendre polynomial expansion of $\rhoB^p$ is
well-behaved and finite as $r\rightarrow \ro$, but the product $\Delta^q
\rhoB^p$ vanishes in the limit $r\rightarrow\ro$ and gives no contribution to
the regularization parameters.

If $q=0$ and $p \ge 2$ and is even, then the $\Theta$ dependence is
in the form of a polynomial in $\cos\Theta$ which has an expansion in
terms of the Legendre polynomials only up to $P_p(\Theta)$, and
because $\rhoB = 0$ when $r=\ro$ and $\Theta=0$ the sum of this
finite number of terms is zero and gives no contribution to the
regularization parameters. This case always results when the
improvement to $\partial_r\psi^\SS$ is $O(\Epsi^n)$ for $n$ being
even.

The only remaining case is $q=0$ and $p > 2$ being a positive odd integer.
In the limit that $r\rightarrow0$,  $\rhoB^p \propto
(1-\cos\Theta)^{p/2}$. The Legendre polynomial expansion of this function
is discussed in detail in Appendix \ref{MathLegendre}. We see in
\Deqn{appAk+1/2} that, for $k$ a positive integer
\beq
  (1-\cos\Theta)^{k+1/2}
       = \sum_{\l=0}^{\infty} \calA^{k+1/2}_\l P_\l(\cos\Theta)
\eeq
where
\beq
  \calA^{k+1/2}_\l = (2\l+1)\calP_{k+1/2}
       /\left[ (2\l-2k-1)(2\l-2k+1)\ldots(2\l+2k+1)(2\l+2k+3) \right],
\label{Ak+1/2}
\eeq
for a constant $\calP_{k+1/2}$ given in \Deqn{Bk+1/2}.  When
$\Theta=0$, $P_\l(\cos\Theta)=1$ and such terms do contribute
additional regularization parameters in the mode sum representation
of the self-force. This case always results when the improvement to
$\partial_r\psi^\SS$ is $O(\Epsi^n)$ for $n$ being odd.

We now see that every other higher order correction to $\psi^\SS$ provides an
additional regularization parameter.  The $\l$ dependence is necessarily of the
form
\beq
  E^k_a \calA^{k+1/2}_\l
\label{Ek/(2l-1)}
\eeq
where $E^k_a$ is independent of $\l$ but still undetermined.  The first term of
this sort, for $k=1$ is included in \Deqn{FABDE}.  It important to note that
for each value of $k$ the sum of these terms from $\l=0$ to infinity is
necessarily zero, and need not be included in the self-force analysis. However
we see in the next section that including these additional coefficients
dramatically speeds up the convergence of the self-force sum \Deqn{sumoverl}.

\section{Application}
\label{Fitting}

In this section we apply the formalism developed above to determine the
self-force $\calF^\R_r$ for a scalar charge in orbit at $r=10M$ about a
Schwarzschild black hole. In the numerical work we use units where $M=1$
and $q=1$.  Appendix \ref{Numerical} describes the practical details for
numerically integrating the scalar wave equation to determine the
$\calF^\ret_{\l r}$. To compute the self-force, the $A_r$, $B_r$ and $D_r$
terms must first be removed from $\calF^\ret_{\l r}$ as in
\Deqns{sumoverl} and (\ref{FABDE})
--- a process which determines residuals which fall off as $\l^{-4}$ for
large $\l$. Removing the contribution of each successive $E^k_r$
improves the falloff of the residuals by an additional two powers of
$\l$.

If we have $\calF^\ret_{\l r}$ for all $\l$, summing the residuals after
removing $A_r$, $B_r$ and $D_r$  would give us the self-force. With
$\calF^\ret_{\l r}$ evaluated only for finite $\l$, we must make some attempt
to obtain the higher $\l$ contributions. To do this we will numerically
determine the $E^k_r$ coefficients by fitting the residuals, considered as a
function of $\l$, with a linear combination of terms whose $\l$ dependence is
given by the $E^1_r$ term in \Deqn{FABDE} for $k=1$ and by \Deqn{Ek/(2l-1)} for
integers $k>1$.

A comparison of the integration results, $\calF^\ret_{\l r}$, for different
values of an accuracy parameter in the numerical routine, revealed that a
systematic effect remained in our best data for $\calF^\ret_{\l r}$. In order
to avoid fitting to that systematic effect, we chose to add to our data a small
random component which was capable of swamping any trace of the systematic
effect, and which allowed us to have precise control of the error in the
$\calF^\ret_{\l r}$. This random component also provided the opportunity to use
Monte Carlo analysis to determine the statistical significance of our result of
the self-force.

\begin{figure}[htbp]
  \centering
  \includegraphics[width=14cm,angle=0]{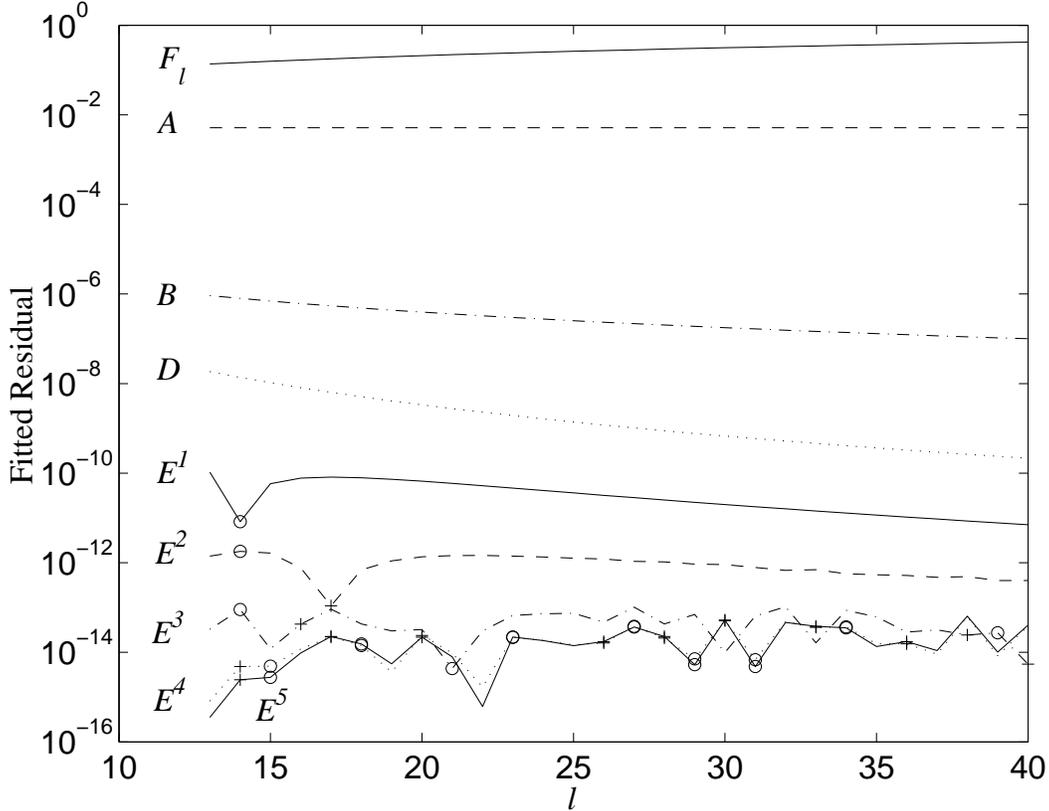} 
  \caption{The upper portion of the figure displays $\calF^\ret_{\l r}$
   as a function of $\l$,
   along with the result of it being regularized by $A_r$, $B_r$, and $D_r$.
   The lower portion displays the residual after a numerical fit
   of from 1 to 5 additional parameters, $E^1_r \ldots E^5_r$.
   A point where the data on a particular curve changes sign from
   being negative to positive is labeled with $+$, from positive to negative by $\circ$.}
  \label{fig1}
\end{figure}

In fitting for the $E^k_r$ we avoided small values of $\l$, which may
contain significant physical information not associated with the large
$\l$ falloff of $\calF^\ret_{\l r}$. Thus we fitted the residues for $\l$
from $13$ to $40$ while determining from $1$ to $5$ of the $E^k_r$
coefficients. Fig.\ \ref{fig1} summarizes the results of this numerical
analysis. The curve labeled $F_\l$ is $\calF^\ret_{\l r}$ as a function of
$\l$. The curves \textit{A}, \textit{B} and \textit{D} show
$\calF^\ret_{\l r} - \calF^\SS_{\l r}$ where $\calF^\SS_{\l r}$
successively includes the contribution from the regularization parameters
$A_r$, $B_r$ and $D_r$. The $E^1$ to $E^5$ curves show the residuals after
numerically fitting from $1$ to $5$ of the $E^k_r$ coefficients and
removing their contributions successively.

\begin{center}
\begin{table}
 \caption{The fitted parameters of $\calF^\ret_{\l r}$} 
 \label{table1}
   \begin{tabular}{l r l l}
     $k$  &  \hskip75pt & $E^k_r$                 & $E^k_r$ part of $\calF^\R_r$ \\
     $1$  &    {$1.$}&{$\!80504(4)\times10^{-4}$} & $2.78201(7)\times10^{-9} $   \\
     $2$  &   {$-1.$}&{$\!000(3)\times10^{-4}$}   & $6.90(2)\times10^{-12}$      \\
     $3$  &    {$4.$}&{$\!3(3)\times10^{-5}$}     & $3.1(2)\times10^{-14}$       \\
     $4$  &   {$-5.$}&{$\!6(5)\times10^{-5}$}     & $7.7(6)\times10^{-16}$
  \end{tabular}
\end{table}
\end{center}

  We actually used a singular value decomposition from Ref.
\cite{NumericalRecipes} to fit the residuals. It provided an independent
estimate of the uncertainty of the $E^k_r$'s, which is entirely compatible
with the Monte Carlo results.  This represented a valuable, overall
consistency check of our analysis. The $E^k_r$ coefficients which result
from a fit of four coefficients are given in Table \ref{table1} along with
their uncertainties in brackets. After fitting four coefficients and
removing their contribution, we obtained an RMS residual of
$2.8\times10^{-14}$ over the fitting range, which is completely determined
by the size of the random component we had introduced to the original
$\calF^\ret_{\l r}$ to swamp the systematic effect. Four coefficients
evidently fit the data down to the noise. It was clear that fitting a
fifth coefficient, or more, did not improve the quality of the fit.

The self-force  $\calF^\R_r = 1.37844828(2)\times 10^{-5}$ was obtained by
summing, over the range of our data, $\calF_{\l r}^\ret$ with the $A_r$,
$B_r$ and $D_r$ terms removed as in \Deqns{sumoverl} and (\ref{FABDE}).
The remainder of the sum to $\l=\infty$ was approximated by the
contributions of the $E^1_r$, $E^2_r$, $\ldots$ sums from 41 to $\infty$,
once the $E^k_r$ coefficients had been determined. The uncertainty was
obtained from the Monte Carlo simulation. The table also shows the
individual contribution of each $E^k_r$ to the self-force $\calF^\R_r$ as
well as the amount of the uncertainty in $\calF^\R_r$ which is
attributable to that $E^k_r$. Without including the effects of the $E^k_r$
tails, we would have found the sum out to $\l=40$ to be $1.37817\times
10^{-5}$. Fitting the higher order terms has allowed us to increase
dramatically the effective convergence to our final result.

Our result is consistent with Fig.~(4A) of Burko's analysis\cite{Burko00}.
With the $A_r$ and $B_r$ terms removed from $\calF_{\l r}^\ret$, he
effectively calculates the total self-force by summing data points on the
equivalent of our curve $B$ out to a large enough value of $\l$ that
convergence is obtained while using Richardson extrapolation.

A future manuscript will apply our methods to the investigation of
physical questions.

\section{Discussion}

In a previous paper \cite{DetWhiting02}, we outlined our method for
computing the self-force. This hinged on realizing that $\psi^\R =
\psi^\ret - \psi^\SS$ is a homogeneous solution of the field equation and
we proved that it gave the same result as methods based on using the tail
part of the retarded Green's function. As a consequence of this we have
obtained the same regularization parameters as all previous authors
\cite{BarackOri00,BarackOri02,BMNOS02,Mino02} using a regularization
procedure based on mode sum expansions. Exact computation of $\psi^\R$
would yield a homogeneous solution of the field equation. Under
interesting physical circumstances, we anticipate that $\psi^\R$ should
consequently have a high level of differentiability \footnote{The
differentiability of $\psi^\R$ is controlled by boundary conditions and
initial data. We consider nondifferentiable initial data or shock waves
coming in from boundaries to be physically unreasonable.}. The level of
differentiability of an approximation for $\psi^\R$ is limited by the
accuracy of the approximation for $\psi^\SS$. To improve the level of
differentiability of our approximation for $\psi^\R$ beyond that of
$\psi^\tail$, we have thus been led to explore higher order approximations
to $\psi^\SS$.

Following earlier work \cite{Det01}, we have used THZ coordinates to
obtain the simple approximation $\psi^\SS\approx q/\rho$. Our key
analytical result is the expansion in \Deqn{drhoinv} which is based on
this approximation. The regularization parameters are derived from the
mode sum representation of each term in \Deqn{drhoinv}. The parameters
$A_r$ and $B_r$ come from the first two terms. The parameter $C_r$ is seen
to be zero, directly from the third term. All these results are consistent
with previous work by others
\cite{BarackOri00,BarackOri02,BMNOS02,Mino02}. Our $D_r$ parameter comes
from the fourth term in \Deqn{drhoinv}, which we compute analytically and
for which we find the specific $\l$ dependence of $1/(2\l-1)(2\l+3)$.
Direct inspection of this term shows that the sum goes to zero in the
coincidence limit, so it does not contribute to the self-force.
Nevertheless, it is recognition of the large $\l$-behavior of the mode sum
expansion for this, and similar higher order terms characterized by the
$E^k_r$, which leads to dramatically improved convergence in the mode
summation. Understanding of the specific nature of the $\l$ dependence in
the mode sum representation of the higher order terms was obtained by an
analysis of general methods for improving the approximation to $\psi^\SS$.
Our numerical application of this scheme amply illustrates the benefits of
estimating the $E^k_r$ parameters in order to accelerate convergence.

In principle, neither the use of $\psi^\R$ instead of $\psi^\tail$, nor
the specific use of THZ coordinates are intrinsically necessary in the
computation of the self-force.  Indeed, many authors have used neither,
and have yet obtained analytical results for the regularization
parameters, and/or numerical results for the self-force
\cite{BarackOri00,BarackOri02,BarackOri03,BMNOS02,Mino02,Lousto00,Lousto01,BarackLousto02}.
It is clear that other methods might also be used to calculate $D_r$ as
well as the general $\l$ dependence of our $E^k_r$ terms. What we believe
is important is that the form of these higher order terms has been
determined, and that the inclusion of these terms has a dramatic impact on
the effectiveness and accuracy of numerical work.

Unquestionably, the intensity of work in this area is paying off, and
as efficient computational techniques are recognized and implemented,
a greater volume of results will become available. This will be
especially true in relation to the long awaited detection of
gravitational waves from binary inspiral sources, represented by a
small compact object in orbit about a comparatively large black hole.

 \acknowledgments
The analysis presented in Appendix \ref{FindTHZcoords} was performed by
Dong Hoon Kim, and we are grateful to Lior Burko for providing us with
some unpublished details of his numerical analysis.  This research has
been supported in part by the Institute of Fundamental Theory at the
University of Florida (E.M),  NSF Grant No. PHY-9800977 (B.F.W.) and NASA
Grant No. NAGW-4864 (S.D.) with the University of Florida. One of us
(B.F.W.) is also grateful to Petr H\'aj\'{\i}\v{c}ek and the Swiss
National Fonds for the opportunity to visit and work at the University of
Bern, Switzerland.

\appendix
\section{The determination of THZ coordinates}
  \label{FindTHZcoords}

A particular THZ coordinate system $(t,x,y,z)$ is associated with any
given geodesic $\Gamma(\tau)$ of a vacuum spacetime and is a harmonic
coordinate system as well as being ``locally inertial and Cartesian.'' By
Zhang's \cite{Zhang86} definition a ``locally inertial and Cartesian''
(LIC) coordinate system has the spatial origin $x^i=0$ on the worldline
$\Gamma$, and has the metric being expandable about $\Gamma$ in powers of
$\rho$ in a particular form which we describe as $g_{ab} = \eta_{ab} +
\rho^p\times$ homogeneous polynomials in $x^i$ of degree $q$, for
non-negative integers $p$ and $q$ with $p+q\ge2$.

The defining features of $n$th order THZ coordinates are that
\begin{enumerate}
  \item[(\textit{i})] On $\Gamma$: $t$ measures the proper time along the geodesic, the
spatial coordinates  $x$, $y$ and $z$ are all zero, $g_{ab}|_\Gamma =
\eta_{ab}$ and all of the first derivatives of $g_{ab}$ vanish.
  \item[(\textit{ii})] At linear, stationary order (\textit{cf.}
Ref.~\cite{Zhang86}) $\HB^{ij} = O(x^{n+1})$.
  \item[(\textit{iii})] The coordinates satisfy the harmonic gauge condition
$\partial_a\gotg^{ab} = O(x^{n})$.
\end{enumerate}
To find THZ coordinates associated with a particular geodesic, it is
easiest to satisfy these conditions in order.

Given a general set of coordinates $Y^A$ and a particular point $p$ a new
set of coordinates $X^a$ may be defined by
\bea
  X^a &=& A^a + B^a{}_A (Y^A-Y_p^A)
\nonumber\\ &&
   + \frac{1}{2} B^a_A\Gamma^A{}_{BC} (Y^B-Y_p^B)(Y^C-Y_p^C) + O[(Y-Y_p)^3]
       \quad Y^A\rightarrow Y_p^A,
\eea
where $\Gamma^A{}_{BC}$ are the usual Christoffel symbols, the $A^a$ are
arbitrary constants, and the $B^a{}_A$ are also arbitrary constants
restricted by the condition that $B^a{}_A$, considered a matrix, be
invertible. Weinberg \cite{Weinberg} shows that
\bea
  g^{ab} &=& g^{AB}\frac{\partial X^a}{\partial Y^A}
                 \frac{\partial X^b}{\partial Y^B}
\nonumber\\ &=&
       \eta^{ab} + O[(Y-Y_p)^2], \quad Y^A\rightarrow Y_p^A,
\eea
so that
\beq
  \frac{\partial g^{ab}}{\partial X^c} =  O[(Y-Y_p)],
                   \quad Y^A\rightarrow Y_p^A.
\eeq
A specific choice for the $A^a$ and $B^a{}_A$ result in the coordinates of
$p$ being $X_p^a = \tau_p \delta^a_t$.  And condition $(i)$ is satisfied
by repeating this construction along $\Gamma$ while parallel propagating
the coordinate basis.

Thus, the coordinates that satisfy condition $(i)$ are denoted $X^a$. And
\bea
  \gotg^{ab} &=& \eta^{ab} - \HB^{ab}
\nonumber\\ &=&
     \eta^{ab} - \HB^{ab}{}_{ij} X^i X^j + O(X^3/\calR^3),
          \quad X^i\rightarrow 0,
\eea
here $X$ in the order term can refer to any of the spatial coordinates,
and the
\beq
  \HB^{ab}{}_{ij} \equiv \left.
        \frac{1}{2}\frac{\partial^2 \HB^{ab}}
              {\partial X^i \partial X^j}\right|_\Gamma
\label{HBabij}
\eeq
are functions only of $t$.

To satisfy conditions $(ii)$ and $(iii)$ we use a gauge transformation of
the form
\beq
   x_{(\new)}^a = X_{(\old)}^a + \zeta^a
\eeq
where changes in geometrical objects are calculated only through linear
order in $\zeta^a$.  In this application, we are interested in the
vicinity of $\Gamma$ and accordingly let
\beq
 \zeta^a =\zeta^a{}_{ijk} X^i X^j X^k ,
\eeq
where the $\zeta^a{}_{ijk}$ are functions only of $t$ to be determined
below. Thus
\beq
  x^a_{(\new)} =  X^a + \zeta^a{}_{ijk} X^i X^j X^k ,
          \quad X^i\rightarrow 0
\label{gaugexnew}
\eeq
changes $\HB^{ab}$ to
\bea
  \HB^{ab}_\new &=& \HB^{ab}_\old + \partial^a \zeta^b + \partial^b \zeta^a
                        - \eta^{ab} \partial_c \zeta^c + O(\zeta^2) .
\label{gaugeHbarnew}
\eea
or
\bea
  \frac13\HB^{abij}_\new &=& \frac13\HB^{abij}_\old + \zeta^{abij} + \zeta^{baij}
                        - \eta^{ab} \zeta_k{}^{kij}.
\eea

We find the necessary gauge transformation in two steps. To satisfy
condition $(ii)$ we require that the gauge transformation obey
\beq
 \partial^i \zeta^j +\partial^j \zeta^i
     - \eta^{ij} \partial_c \zeta^c = - \HB^{ij}_\old + O(X^3),
\label{dzeta+dzeta}
\eeq
which implies that
\beq
   \zeta_{ijkl} + \zeta_{jikl} - \eta_{ij} \zeta^m{}_{mkl}
         = -\frac{1}{3}\HB^\old_{ijkl}.
\label{HBij}
\eeq
  We use the decomposition of STF tensors of reference
\cite{DamourIyer91b} and let
\beq
\zeta_{ijkl} = A_{\langle ijkl\rangle } + \frac{3}{4}\epsilon_{ip\langle
j}B^p{}_{kl\rangle }
                + \frac{5}{7}\delta_{i\langle j} C_{kl\rangle }
                - \frac{1}{5}\HB_i{}^{p}{}_{p(j}\delta_{kl)},
\label{zetaiABC}
\eeq
where $\langle \ldots\rangle $ implies taking the STF part of the enclosed
indices and $A_{ijkl}$, $B_{pkl}$ and $C_{kl}$ are STF tensors on all of
their indices. The coefficient $-1/5$ in the last term is chosen to make the
trace of \Deqn{zetaiABC} agree with the trace of \Deqn{HBij}. The remainder
of the solution for $\zeta_{ijkl}$ is
\beq
  A_{ijkl} = - \frac{1}{6} \HB_{\langle ijkl\rangle }
             + \frac{1}{7} \delta_{(ij} \HB^p{}_{kl)p}
             + \frac{1}{42} \HB^p{}_{p(ij}\delta_{kl)}
             - \frac{1}{35} \delta_{(ij}\delta_{jk)} \HB^{pq}{}_{pq},
\eeq
\beq
  B_{pkl} = - \frac{1}{3} \epsilon_{ij(p} \HB^i{}_{kl)}{}^j
            + \frac{1}{15} \delta_{(kl}
               \epsilon_{p)}{}^{ij} \HB_{iq}{}^q{}_j
\eeq
and
\beq
  C_{kl} = \frac{1}{3} \HB^i{}_{ikl}
          + \frac{1}{15} \left(\HB^p{}_{jkp} + \HB^p{}_{kjp}
          + \delta_{jk} \HB^{pq}{}_{pq} \right) ,
\eeq
but only if $\HB_{ijkl}$ satisfies the auxiliary condition that
\beq
  \HB_{kli}{}^i - \HB^i{}_{kli} - \HB^i{}_{lki}
              + \delta_{kl}\HB^{pq}{}_{pq} = 0.
\label{AuxCond}
\eeq
This condition is automatically satisfied when $\HB^{ab}$ is derived from
a vacuum solution of the Einstein equations, cf. Eq.~(35.64) of
Ref.~\cite{MTW}.

The spatial component of condition $(iii)$ is now satisfied through $O(x)$
because $\HB^{ij}$ vanishes. The time component of condition $(iii)$ will
be satisfied through $O(x)$ also if
\beq
  \partial^b \partial_b \zeta^t = - \partial_b \HB^{bt}_\old ,
\eeq
thus
\beq
  \partial^b \partial_b \zeta^t =
     - \ddot{\zeta}^t{}_{ijk}X^i X^j X^k
     + 6 \zeta^{ti}{}_{ik} X^k
     =
       - 2 \HB^{ti}{}_{ik} X^k + O(X^2)
            \quad X^i\rightarrow 0,
\eeq
or
\beq
  \zeta^{ti}{}_{ik}
     = - \frac{1}{3} \HB^{ti}{}_{ik} .
\eeq
An elementary solution for $\zeta^t{}_{ijk}$ is
\beq
  \zeta^t{}_{ijk} = -\frac{1}{5} \HB^{tp}{}_{p(i} \delta_{jk)}.
\label{zetaa}
\eeq

The combination of \Deqns{zetaiABC} and (\ref{zetaa}) provides the
necessary gauge transformation that results in satisfaction of conditions
$(ii)$ and $(iii)$ for second order THZ coordinates.

The third order coordinates may be found following a similar procedure
where the gauge transformation is of the form
\beq
  x^a_{(\new)} =  x_{(\old)}^a + \zeta^a{}_{ijkl} X^i X^j X^k X^l,
          \quad X^i\rightarrow 0 .
\label{gaugexnew4}
\eeq

At the fourth and higher orders, where terms in the metric expansion
involving two derivatives of the Riemann tensor are of the same order as
terms quadratic in the Riemann tensor, the presence of nonlinearities
complicate the construction of THZ coordinates.  We have not needed
detailed knowledge about these higher order THZ coordinates throughout
this paper.

\section{THZ coordinates for a circular orbit of Schwarzschild}
  \label{ActualTHZ}

To find the THZ coordinates for a circular orbit in the Schwarzschild
geometry it is convenient to take advantage of the spherical symmetry of
the background geometry rather than to follow the general procedure
described in the preceding section.  The orbit $\Gamma$, given by
$\phi=\Omega t$  where $\Omega = \sqrt{M/\ro^3}$ is the orbital frequency
at Schwarzschild radius $\ro$, is tangent to a Killing vector field $\xi^a
\equiv \partial/ \partial t_s + \Omega
\partial /\partial \phi$. We choose three helping functions $\calX$,
$\calY$ and $\calZ$ which are Lie derived by the Killing vector,
$\Lie_\xi\calX = \Lie_\xi\calY = \Lie_\xi\calZ = 0$, and their
gradients are spatial and orthogonal to the 4-velocity of the
geodesic and to each other when evaluated on $\Gamma$. These helping
functions are
\beq
 \XX \equiv \frac{r-\ro}{(1-2M/\ro)^{1/2}},
\label{calX}
\eeq
\beq
  \YY \equiv \ro \sin\theta \sin(\phi-\Omega t_\s)
                   \left(\frac{\ro-2M}{\ro-3M}\right)^{1/2}
\label{calY}
\eeq
and
\beq
  \ZZ \equiv  \ro \cos(\theta).
\label{calZ}
\eeq

Two more useful functions are
\bea
   \xx &=& \frac{ [r \sin\theta \cos(\phi-\Omega t_\s)-\ro] }{(1-2M/\ro)^{1/2}}
\nonumber \\ &&
  {} + \frac{ \M}{\ro^2 (1-2M/\ro)^{1/2}}
         \left[ - \frac{\XX^2}{2} + \YY^2
                   \left(\frac{\ro -3 \M}{\ro -2 \M}\right) + \ZZ^2\right]
\nonumber \\ &&
  {} + \frac{ \M \XX}{2 \ro^3 (\ro -2 \M) (\ro -3 \M)}
          \left[- \M^2 \XX^2 +  \YY^2(\ro -3 \M)(3\ro-8\M)
                + 3 \ZZ^2(\ro -2 \M)^2 \right]
\nonumber \\ &&
  {} + \frac{ \M}{\ro^5(1-2M/\ro)^{1/2}(\ro -3 \M)} \left[
  \M \XX^4 \frac{(\ro^2 - \ro  \M + 3 \M^2)}{8 (\ro -2 \M)} \right.
\nonumber \\ &&
  {} + \frac{\XX^2 \YY^2}{28} (28 \ro^2 - 114 \ro \M + 123 \M^2)
   + \frac{\XX^2 \ZZ^2}{14} (14 \ro^2 -  48 \ro \M +  33 \M^2)
\nonumber \\ &&
  {} + \frac{\M\YY^4}{56 (\ro -2 \M)^2} \left(3 \ro^3 - 74 \ro^2 \M + 337 \ro \M^2
                - 430 \M^3\right)
\nonumber \\ &&
  \left. {}
       - \frac{\M^2 \YY^2 \ZZ^2 (7 \ro -18 \M)}{4 (\ro -2 \M)}
       - \frac{\M \ZZ^4}{56} (3\ro +22 \M) \right]
\label{thzxbar}
\eea
and
\bea
 \yy &=&  r \sin\theta \sin(\phi-\Omega t_\s)
                              \left(\frac{\ro-2M}{\ro-3M}\right)^{1/2}
  {} + \frac{ \M \YY}{2 \ro^3}
      \left[ - 2 \XX^2
            + \YY^2 \left(\frac{\ro -3 \M}{\ro -2 \M}\right) + \ZZ^2\right]
\nonumber \\ &&
  {} + \frac{  \M  \XX \YY }{14 \ro^5 (1-2M/\ro)^{1/2} (\ro -3 \M)}
            \left[ 2 \M \XX^2 (4 \ro -15 \M) \right.
\nonumber \\ &&
   \left. {} +   \YY^2 (14 \ro^2-69\M\ro +89 \M^2)
          + 2 \ZZ^2 (\ro-2 \M) (7 \ro -24 \M) \right],
\label{thzybar}
\eea

In terms of the functions defined above, the THZ coordinates $(t,x,y,z)$
are
\beq
  \x = \xx \cos(\Omega^\dag t_\s) - \yy \sin(\Omega^\dag t_\s)
\label{thzx}
\eeq
and
\beq
  \y = \xx \sin(\Omega^\dag t_\s) + \yy \cos(\Omega^\dag t_\s)
\label{thzy}
\eeq
where $\Omega^\dag=\Omega\sqrt{1-3 \M/\ro}$, along with
\bea
 \z &=& r \cos(\theta)
   {} + \frac{ \M \ZZ}{2 \ro^3 (\ro -3 \M)}
               \left[ - \XX^2(2 \ro -3 \M)
                    + \YY^2(\ro -3 \M) + \ZZ^2(\ro -2 \M)\right]
\nonumber \\ &&
   {} + \frac{  \M  \XX \ZZ}{14 \ro^5 (1-2M/\ro)^{1/2} (\ro -3 \M)}  \left[
                 \M  \XX^2 (13 \ro -19 \M) \right.
\nonumber \\ &&
    \left. {}  + \YY^2 (14\ro^2 - 36\ro\M + 9\M^2)
            + \ZZ^2 (\ro -2 \M)(14 \ro -15 \M) \right])
\label{thzz}
\eea
and
\bea
  \t &=& t_\s(1-3M/\ro)^{1/2}   - \frac{ r \Om\YY}{(1-2\M/\ro)^{1/2}}
\nonumber \\ &&
   {} + \frac{ \Om \M \YY}{\ro^2 (1-2M/\ro)^{1/2} (\ro -3 \M)}
       \left[ -\frac{\XX^2}{2}(\ro - \M)
           + \M \YY^2 \frac{\ro -3 \M}{3 (\ro -2 \M)} + \M \ZZ^2 \right]
\nonumber \\ &&
   {} + \frac{ \Om \M \XX \YY}{14 \ro^3 (\ro -2 \M) (\ro -3 \M)} \left[
             - \XX^2(\ro^2-11 \ro  \M+11 \M^2) \right.
\nonumber \\ &&
  \left.  + \YY^2 (13\ro^2 - 45\ro\M + 31\M^2)
          + \ZZ^2 (13\ro - 5\M) (\ro-2\M)  \right] .
\label{thzt}
\eea

The set of functions $(t,\xx,\yy,z)$ forms a non-inertial coordinate
system that co-rotates with the particle in the sense that the $\xx$ axis
always lines up the center of the black hole and the center of the
particle, the $\yy$ axis is always tangent to the spatially circular
orbit, and the $z$ axis is always orthogonal to the orbital plane. The
spatial coordinates are all Lie derived by the Killing vector,
$\Lie_\xi\xx = \Lie_\xi\yy = \Lie_\xi z = 0$.

The $x^a$ coordinates $(t,x,y,z)$ are locally inertial and non-rotating in
the vicinity of $\Gamma$, but these same coordinates appear to be rotating
when viewed far from $\Gamma$ as a consequence of Thomas precession as
revealed in the $\Omega^\dag t_\s$ dependence in \Deqns{thzx} and
(\ref{thzy}) above.

The determination of the THZ coordinates was tedious but not difficult. We
looked for the relationship between the THZ coordinates and the usual
Schwarzschild coordinates $X^A=(t_\s,r,\theta,\phi)$ by using the usual
rule for the change in components of a tensor under a coordinate
transformation,
\beq
  g^{ab} = g^{AB} \frac{\partial \x^a}{\partial X^A}
                  \frac{\partial \x^b}{\partial X^B}
\label{THZmetric}
\eeq
where $g^{AB}$ is the Schwarzschild geometry in the Schwarzschild
coordinates. The terms through $O(\calX^2)$ (the $\calX$ in the order term
represents any of $\calX$, $\calY$, or $\calZ$) in the definitions of $t$,
$\xx$, $\yy$ and $z$ and the rotation represented in \Deqns{thzx} and
(\ref{thzy}) were chosen so that $t$ measured the proper time on the
orbit,  $g_{ab}|_\Gamma =\eta_{ab}$ and $\partial_c g_{ab}|_\Gamma=0$.
This much could be done easily by hand and resulted in coordinates that
satisfied condition $(i)$ in Appendix \ref{FindTHZcoords}.

The $O(\calX^3)$ and $O(\calX^4)$ terms were found by use of
\textsc{grTensor} running under \textsc{Maple} following a procedure
similar to that in Appendix \ref{FindTHZcoords} except that we used
homogeneous polynomials of the form $\zeta^a{}_{ijk}\calX^i\calX^j\calX^k$
and $\zeta^a{}_{ijkl}\calX^i\calX^j\calX^k\calX^l$ along with
\Deqn{THZmetric} to determine $\zeta^a{}_{ij\ldots}$ which resulted in the
satisfaction of conditions $(ii)$ and $(iii)$ in Appendix
\ref{FindTHZcoords}.

Note that ultimately the THZ coordinates $(t,x,y,z)$ are linear
combinations of products of $C^\infty$ functions of the Schwarzschild
coordinates, and so are $C^\infty$ functions themselves.

It is convenient to note that the natural Minkowski metric that goes with
the Thorne-Hartle coordinates is $\eta_{ab}\,dx^a \,dx^b \equiv
-d\t^2+d\x^2+d\y^2+d\z^2$, while its components in the original
Schwarzschild coordinates are
\beq
   \eta_{AB} = \eta_{ab} \frac{\partial x^a}{\partial X^A}
       \frac{\partial x^b}{\partial X^B}.
\eeq
Another form for $\eta_{ab}$ is
\beq
  \eta_{ab} = -\nabla_a \t\; \nabla_b \t + \nabla_a\x\; \nabla_b\x
      +\nabla_a\y\; \nabla_b\y
            + \nabla_a\z\; \nabla_b\z
\eeq
or
\bea
  \eta_{ab} &=& -\nabla_a t\; \nabla_b t
            + \nabla_a\xx \;\nabla_b\xx +\nabla_a\yy\; \nabla_b\yy
            + \nabla_a\zz\; \nabla_b\zz
\nonumber\\ &&
      {} + (\xx^2+\yy^2) \nabla_a(\Omega^\dag t_s) \nabla_b(\Omega^\dag t_s)
         + 2[\xx\nabla_{(a}\yy - \yy\nabla_{(a}\xx ] \nabla_{b)}(\Omega^\dag
             t_s).
\eea
And from the above definitions it readily follows that $\Lie_\xi \rho^2 =
\Lie_\xi z = \Lie_\xi \nabla_a t = 0$, while $\Lie_\xi x$, $\Lie_\xi y$,
and $\Lie_\xi t$ are nonzero.

\section{Integrals over $\Phi$}
  \label{MathEllip}

The approach in this and the following Appendix, is similar to that of
Appendices C and D of Ref.~\cite{Mino02}.

 In Section\ref{RegularizationP} we define
\beq
  \chi \equiv 1 - \alpha \sin^2\Phi
\eeq
where
\beq
  \alpha \equiv \frac{M}{\ro-2M} .
\eeq
And we use
\bea
  \left\langle \chi^{-p} \right\rangle
      =  \left\langle (1-\alpha\sin^2\Phi)^{-p} \right\rangle &=&
       \frac{2}{\pi} \int_0^{\pi/2} (1-\alpha\sin^2\Phi)^{-p} \, d\Phi
\nonumber\\
  &=& {}_2F_1(p,\frac{1}{2};1;\alpha)
  \equiv F_p .
  \label{s2ave}
\eea
This result follows almost immediately from
\bea
  \left\langle \chi^{-p} \right\rangle  &=& \frac{2}{\pi}\int_0^{\pi/2}
      (1-\alpha\sin^2\Phi)^{-p} \, d\Phi
\nonumber\\ &=&
  \frac{1}{\pi}\int_0^1 t^{-1/2} (1-t)^{-1/2}
      \left( 1-\alpha t \right)^{-p}\, dt ,
\eea
where $t=\sin^2\Phi$, and from the integral representation of the
hypergeometric function, equation~(15.3.1) of reference~\cite{AandS}
\beq
  {}_2F_1(a,b;c;z) = \frac{\Gamma(c)}{\Gamma(b)\Gamma(c-b)}
    \int_0^1t^{b-1}(1-t)^{c-b-1}(1-tz)^{-a}\,dt ,
    \quad \calR e (c)>\calR e(b)>0.
\eeq

Two elementary special cases of \Deqn{s2ave} are
\beq
  \left\langle 1-\alpha\sin^2\Phi \right\rangle
      = {}_2F_1(-1,\frac12;1;\alpha) = 1-\frac12\alpha = F_{-1}
\eeq
and
\beq
  \left\langle (1-\alpha\sin^2\Phi)^{-1} \right\rangle
      = {}_2F_1(1,\frac12;1;\alpha) = (1-\alpha)^{-1/2} = F_1 ;
\label{F1}
\eeq
The latter is used in \Deqn{firstterm} and leads to the $A_r$ term in
\Deqn{FABDE}. The special cases $p=\frac12$ and $p=-\frac12$ are also
easily represented in terms of complete elliptic integrals of the first
and second kinds respectively,
\beq
  \frac{2}{\pi} K(\alpha) = {}_2F_1(\frac12,\frac12,1;\alpha)
                        = F_{1/2}
\eeq
and
\beq
  \frac{2}{\pi} E(\alpha) = {}_2F_1(-\frac12,\frac12,1;\alpha)
                        = F_{-1/2} .
\eeq

\section{Legendre polynomial expansions}
  \label{MathLegendre}

We require the coefficients $\calA^{p/2}_\l (\delta)$ in the expansion
\beq
 \left(\delta^2 + 1 - u \right)^{p/2}
       = \sum_{\l =0}^\infty \calA^p_\l (\delta) P_\l (u ),
       \quad \text{for } \delta\rightarrow0 ,
\label{gendecomp}
\eeq
for both positive and negative odd-integral values of $p$. Note that if $p$
is a positive even integer then the left hand side is a $p/2$ degree
polynomial in $u$ and the sum terminates with $\l =p$.

First we analyze the negative odd-integral values of $p$ via induction.
 The generating function for Legendre polynomials is
\beq
   \left(1-2tu +t^2\right)^{-1/2} =
         \sum_{\l =0}^\infty t^\l  P_\l (u ),  \quad |t|<1 .
\label{Plgen}
\eeq
With $T$ defined from
\beq
  t = e^{-T},
\eeq
\Deqn{Plgen} implies
\beq
   \left(e^T+e^{-T} -2u \right)^{-1/2} =
         \sum_{\l =0}^\infty e^{-(\l+1/2)T}  P_\l (u),  \quad T> 0 .
\label{Pleg-1/2}
\eeq
The expansion
\beq
  e^T + e^{-T} = 2 + T^2 + O(T^4), \quad T \rightarrow 0,
\label{expanT}
\eeq
followed by the substitution
\beq
  T = \delta \sqrt{2}
\label{subD}
\eeq
in \Deqn{Pleg-1/2} provides
\beq
  \calA^{-1/2}_\l  = \sqrt{2}+O(\l \delta ), \quad \delta \rightarrow 0,
\label{Al-1/2}
\eeq
which is used in the second term of \Deqn{drhoinv} to arrive at
\Deqn{secondterm} and leads to the $B_r$ term of \Deqn{FABDE} and to the
absence of a term in \Deqn{FABDE} which might have resulted from the third
term of \Deqn{drhoinv}.

Differentiation of both sides of \Deqn{Pleg-1/2} with respect to $T$ yields
\beq
   -\frac{1}{2}\left(e^T+e^{-T} -2u \right)^{-3/2} \left(e^T-e^{-T}\right)
      =  \sum_{\l =0}^\infty - \left(\l+\frac{1}{2}\right) e^{-(\l+1/2)T}  P_\l (u),
      \quad T> 0 .
\eeq
Simplification and expansion about $T=0$ gives
\beq
   \left(e^T+e^{-T} -2u \right)^{-3/2}
      =  \sum_{\l =0}^\infty \frac{\left(2\l +1\right)}{2T}  P_\l (u)
         \left[ 1 + O(\l T)\right],
      \quad T\rightarrow 0,
\eeq
and repeated differentiation extends this result to
\beq
   \left(e^T+e^{-T} -2u \right)^{-k-1/2}
      =  \sum_{\l =0}^\infty \frac{\left(2\l +1\right)}{2(2k-1)T^{2k-1}}  P_\l (u)
         \left[ 1 + O(\l T)\right],
      \quad T\rightarrow 0.
\eeq
Finally, for $k \ge 1$ the expansion and substitution of \Deqns{expanT}
and (\ref{subD}) result in
\beq
  \calA_\l ^{-k-1/2} = \frac{2\l +1}{\delta^{2k-1}(2k-1)}[1+O(\l \delta)]
        \quad \delta \rightarrow 0 .
\label{A-k-1/2}
\eeq
For $k=1$
\beq
  \calA^{-3/2}_\l  =
    \frac{2\l +1}{\delta } [1+O(\l \delta )],
  \quad \delta\rightarrow 0,
  \label{Al-3/2}
\eeq
is used in the first term of \Deqn{drhoinv} to obtain \Deqn{firstterm} and,
subsequently, the $A_r$ term of \Deqn{FABDE}.

Next, for positive odd-integral values of $p$ in \Deqn{gendecomp}, first let
$p=1$ and multiply the left hand side of \Deqn{Plgen} by
 $(1-u)^{1/2}$ and the right hand side by $\sum_\l A_\l^{1/2}P_\l(u)$.
Then, integrate over $u$ from $-1$ to $1$; the right hand side is
\beq
    \sum_\l t^\l A_\l^{1/2}2/(2\l + 1)
\eeq
from the normalization of the Legendre polynomials,
\beq
  \int_{-1}^1 P_\l (u ) P_{\l ^\prime}(u ) \,du
     = \frac{2\delta_{\l \l ^\prime}}{2\l +1}.
\eeq
Now, expand the left hand side in powers of $t$ to determine the
$\calA^{1/2}_\l $. This results in
\beq
   (1-u )^{1/2} = \sum_{\l =0}^\infty
         \frac{-2\sqrt{2}}{(2\l -1)(2\l +3)} P_\l (u )
\label{(1-u)1/2}
\eeq
and
\beq
  \calA^{1/2}_\l  = \frac{-2\sqrt{2}}{(2\l -1)(2\l +3)}.
\label{Al+1/2}
\eeq
The latter is used in the $\Epsi^1$ term of \Deqn{drhoinv} to obtain
\Deqn{lastterm} and, subsequently, the $D_r$ term of \Deqn{FABDE}.

For other positive odd-integral values of $p>1$ in \Deqn{gendecomp},
consider the Legendre polynomial representation of $(1-u )^{k+1/2}$, with
$k$ a positive integer,
\beq
  (1-u )^{k+1/2} = \sum_{\l =0}^\infty \calA^{k+1/2}_\l  P_\l (u )
\label{legexpand}
\eeq
which defines the expansion coefficients $\calA^{k+1/2}_\l $.
 The first coefficient $\calA^{k+1/2}_0$ is obtained by multiplying both sides
of \Deqn{legexpand} by $1=P_0(u)$, integrating over $u$ from $-1$ to $1$
and using the orthogonality of the Legendre polynomials to yield
\beq
  \calA^{k+1/2}_0 = \frac{2^{k+1/2}}{k+\frac{3}{2}} .
\label{A0k+1/2}
\eeq

The coefficients $\calA^{k+1/2}_l$ for $\l\ge1$ are obtained from
\Deqn{(1-u)1/2} by induction on $k$.
 The derivative of \Deqn{legexpand} provides
\bea
  \sum_{\l =0}^\infty \calA^{k+1/2}_\l  P^\prime_\l    &=&
     - \left(k+\frac12\right) (1-u )^{k-1/2}
\nonumber\\ &=&
    -\left(k+\frac12\right) \sum_{\l =0}^\infty \calA^{k-1/2}_\l  P_\l
\nonumber\\ &=&
 \left(k+\frac12\right) \sum_{\l =0}^\infty
   \calA^{k-1/2}_\l  \frac{P^\prime_{\l -1} - P^\prime_{\l +1}}{2\l +1} ,
\eea
where the prime denotes differentiation with respect to $u $, and the last
equality follows from equation~(12.23) of reference \cite{Arfken}. A
re-summation of this last expression yields
\beq
  \sum_{\l =1}^\infty \calA^{k+1/2}_\l  P^\prime_\l  =
        \left(k+\frac12\right) \sum_{\l =1}^\infty
            \left[ \frac{\calA^{k-1/2}_{\l +1}}{2\l +3}
               - \frac{ \calA^{k-1/2}_{\l -1}}{2\l -1} \right]  P^\prime_{\l } .
\eeq
For $\l\ge 1$
\beq
 \calA^{k+1/2}_\l   =
     \left(k+\frac12\right)
         \left[ \frac{\calA^{k-1/2}_{\l +1}}{2\l +3}
               - \frac{ \calA^{k-1/2}_{\l -1}}{2\l -1}\right]
\label{inductOnM}
\eeq
provides $A_\l^{k+1/2}$ in terms of $A_\l^{k-1/2}$ with the help of
\Deqn{A0k+1/2}. The final result is
\beq
  \calA^{k+1/2}_\l  = \calP_{k+1/2}(2\l +1)
       /\left[ (2\l -2k-1)(2\l -2k+1)\ldots(2\l +2k+1)(2\l +2k+3) \right],
\label{appAk+1/2}
\eeq
where
\beq
   \calP_{k+1/2} = (-1)^{k+1} 2^{k+3/2}\left[(2k+1)!!\right]^2 .
\label{Bk+1/2}
\eeq
for $k$ a positive integer or zero. \Deqn{appAk+1/2} is used to furnish the
$\l$ dependence in the $E^1_r$ term of \Deqn{FABDE}.

The significant conclusions of this Appendix are summarized in
\Deqns{Al-1/2}, (\ref{A-k-1/2}) and (\ref{appAk+1/2}).

\section{Integration of the scalar wave equation}
\label{Numerical}

The scalar field resulting from a charge $q$ moving in a circular orbit of
the Schwarzschild geometry is most easily found following an approach
similar to that of Breuer {\it et al.} \cite{Breuer73} or, more recently,
Burko \cite{Burko00}. The wave equation for the scalar field is
\beq
   \nabla^2\psi = -4\pi\varrho,
\label{DelPsi2}
\eeq
where the scalar field source $\varrho$, being distinct from $\rho$,
represents a point charge $q$ moving through spacetime along a worldline
$\Gamma(\tau)$, described by coordinates $z^a(\tau)$. This source is
\bea
   \varrho(x) &=& q \int (-g)^{-1/2} \delta^4(x^a-z^a(\tau)) d\tau
\nonumber\\
           &=& q (-g)^{-1/2}(dt/d\tau)^{-1} \delta^3(x^i-z^i(t)) ,
\label{qpp2}
\eea
with $\tau$ the proper time along the worldline. For a circular orbit at
radius $\ro$, expanding $\varrho$ in terms of spherical harmonic
components provides
\bea
   \varrho & = & q \int  (-g)^{-1/2} \delta(r-\ro) \delta(\theta-\pi/2)
                       \delta(\phi-\Omega t) \delta(t-t(\tau)) d\tau
\nonumber\\
        & = & r^{-2} q \delta(r-\ro)\delta(\theta-\pi/2)\delta(\phi-\Omega t)
   dt/d\tau
\nonumber\\  & = &\sum_\lm \frac{q_\lm}{4\pi \ro}
            \delta(r-\ro)e^{i\omega_m t} Y_\lm(\theta,\phi) ,
\eea
where
\beq
  \omega_m \equiv -m\Omega,
\eeq
\begin{equation}
  q_{\ell m} = \frac{4\pi q}{\ro} \frac{Y^\ast_{\ell
      m}(\pi/2,0)}{dt/d\tau},
\end{equation}
and
\beq
 \frac{dt}{d\tau} = \frac{1}{\sqrt{1-3M/\ro}}.
\eeq

Also, decomposing $\psi$ provides
\beq
   \psi =\sum_{\l,m} \psi_{\l m}(r) e^{i\omega_mt} Y_{\lm}(\theta,\phi),
\eeq
and the $\l m$ component of the scalar wave equation becomes
\begin{equation}
  \frac{d^2\psi_{\ell m}}{dr^2} + \frac{2(r-M)}{r(r-2M)} \frac{d\psi_{\ell m}}{dr}
    + \left[ \frac{\omega^2 r^2}{(r-2M)^2}  - \frac{\ell(\ell+1)}{r(r-2M)} \right]
      \psi_{\ell m}
    = - \frac{q_{\ell m}}{\ro-2M} \delta(r-\ro) .
\label{d2psi}
\end{equation}

We know that
\beq
  Y_{\l,-m} = (-1)^m Y^\ast_{\l,m} ,
\eeq
and the reality of $\varrho$ and of the final solution for
$\psi(t,r,\theta,\phi)$ requires similar expressions for $ q_{\l,-m}$ and
$\psi_{\l,-m}$.

The boundary conditions of interest require only ingoing waves at the
event horizon
\begin{equation}
  \psi_{\ell m} = e^{i\omega r_\ast}/r, \quad r\rightarrow2M,
\label{ehBC}
\end{equation}
and only outgoing waves at infinity
\begin{equation}
  \psi_{\ell m} = e^{-i\omega r_\ast}/r \quad r\rightarrow\infty,
\label{inftyBC}
\end{equation}
where
\beq
  r_\ast=r+2M\log(r/2M -1) .
\eeq

An expansion of $\psi_\lm$ starts the numerical integration at large
$r$. We assume that
\beq
  \psi_\lm(r) = \frac{e^{-i\omega r_\ast}}{r} \sum_{n=0} \frac{a_n}{r^n}
\label{expanInfty}
\eeq
and, with \Deqn{d2psi}, obtain a recursion relation for $a_n$:
\beq
  a_n = \frac{n(n-1) - \l(\l+1)}{2i\omega n} a_{n-1}
             - \frac{M(n-1)^2}{i\omega n} a_{n-2},
\label{rrInfty}
\eeq
with the starting values of $a_0=1$ and $a_{n<0} = 0$. This is an asymptotic
expansion, and we begin the integration of \Deqn{d2psi} at a value of $r$ which
is just big enough that the sum in \Deqn{expanInfty} reaches machine accuracy
before beginning to diverge. We numerically integrate \Deqn{d2psi} in to the
radius of the orbit $\ro$: this provides us with a homogeneous solution
$\psi^\infty$ with proper boundary conditions at large $r$.

Similarly an expansion of $\psi_\lm$ for small $r-2M$ starts the numerical
integration near the event horizon.  We assume that
\beq
  \psi_\lm(r) = \frac{e^{i\omega r_\ast}}{r} \sum_{n=0} b_n(r-2M)^n
\label{expanHorizon}
\eeq
and, with \Deqn{d2psi}, obtain a recursion relation for $b_n$:
\bea
  b_n &=& -\frac{12i\omega M(n-1)+(2n-3)(n-1) - (\l^2+\l+1)}{2M(4in\omega M+n^2)} b_{n-1}
\nonumber\\
    && {} - \frac{12i\omega M(n-2)+(n-2)(n-3)-\l(\l+1)}{4M^2(4in\omega M+n^2)}
            b_{n-2}
\nonumber\\
    && {} - \frac{i\omega(n-3)}{2M^2(4in\omega M+n^2)} b_{n-3} ,
\label{rrHorizon}
\eea
with the starting values of $b_0=1$ and $b_{n<0} = 0$.  We begin the
integration of \Deqn{d2psi} at a value of $r-2M$ which is just small
enough that the sum in \Deqn{expanHorizon} reaches machine accuracy
within a reasonable number of terms.  We numerically integrate
\Deqn{d2psi} out to the radius of the orbit $\ro$: this provides us
with a homogeneous solution $\psi^H$ with proper boundary conditions
near the event horizon.

The retarded field is
\beq
  \psi^\ret_\lm = \begin{cases}
      A \, \psi^H_\lm, & r <\ro \\
      B \, \psi^\infty_\lm, & r >\ro ,
  \end{cases}
\eeq
with the match at $\ro$ determined by the $\delta-$function source of
\Deqn{d2psi},
\beq
  \left( B\frac{d\psi^\infty_\lm}{dr} - A\frac{d\psi^H_\lm}{dr} \right)_\ro =
              -\frac{q_\lm}{\ro-2M}
\label{discontinuity}
\eeq
which yields
\beq
  A \left( \psi^H_\lm \frac{d\psi^\infty_\lm}{dr}
               - \psi^\infty_\lm \frac{d\psi^H_\lm}{dr} \right)
  =  - \psi^\infty_\lm \frac{q_\lm}{\ro-2M},
\eeq
and
\beq
  B \left( \psi^H_\lm \frac{d\psi^\infty_\lm}{dr}
               - \psi^\infty_\lm \frac{d\psi^H_\lm}{dr} \right)
  =  - \psi^H_\lm \frac{q_\lm}{\ro-2M} .
\eeq
The $\l$ component of the radial self-force for $\psi^\ret$ in
\Deqn{sumoverl}, is then given by
\beq
 \calF^\ret_{\l r} = \sum_{m=-\l}^\l \left.
    \frac{d\psi^\ret_\lm}{dr}\right|_{\ro} .
\label{Fretlr}
\eeq
Section \ref{Fitting} describes the efficient use of
 $\calF^\ret_{\l r}$ in the determination of the self-force.


\begin{thebibliography}{29}
\expandafter\ifx\csname natexlab\endcsname\relax\def\natexlab#1{#1}\fi
\expandafter\ifx\csname bibnamefont\endcsname\relax
  \def\bibnamefont#1{#1}\fi
\expandafter\ifx\csname bibfnamefont\endcsname\relax
  \def\bibfnamefont#1{#1}\fi
\expandafter\ifx\csname citenamefont\endcsname\relax
  \def\citenamefont#1{#1}\fi
\expandafter\ifx\csname url\endcsname\relax
  \def\url#1{\texttt{#1}}\fi
\expandafter\ifx\csname urlprefix\endcsname\relax\def\urlprefix{URL }\fi
\providecommand{\bibinfo}[2]{#2} \providecommand{\eprint}[2][]{\url{#2}}

\bibitem[{\citenamefont{Detweiler and Whiting}(2003)}]{DetWhiting02}
\bibinfo{author}{\bibfnamefont{S.}~\bibnamefont{Detweiler}} \bibnamefont{and}
  \bibinfo{author}{\bibfnamefont{B.~F.} \bibnamefont{Whiting}},
  \bibinfo{journal}{Phys. Rev. D} \textbf{\bibinfo{volume}{67}},
  \bibinfo{pages}{024025} (\bibinfo{year}{2003}), \eprint{gr-qc/0202086}.

\bibitem[{\citenamefont{Dirac}(1938)}]{Dirac38}
\bibinfo{author}{\bibfnamefont{P.~A.~M.} \bibnamefont{Dirac}},
  \bibinfo{journal}{Proc. R. Soc. (London)} \textbf{\bibinfo{volume}{A167}},
  \bibinfo{pages}{148} (\bibinfo{year}{1938}).

\bibitem[{\citenamefont{De{W}itt and Brehme}(1960)}]{DeWittBrehme60}
\bibinfo{author}{\bibfnamefont{B.~S.} \bibnamefont{De{W}itt}} \bibnamefont{and}
  \bibinfo{author}{\bibfnamefont{R.~W.} \bibnamefont{Brehme}},
  \bibinfo{journal}{Ann. Phys.} \textbf{\bibinfo{volume}{9}},
  \bibinfo{pages}{220} (\bibinfo{year}{1960}).

\bibitem[{\citenamefont{Mino et~al.}(1997)\citenamefont{Mino, Sasaki, and
  Tanaka}}]{Mino97}
\bibinfo{author}{\bibfnamefont{Y.}~\bibnamefont{Mino}},
  \bibinfo{author}{\bibfnamefont{M.}~\bibnamefont{Sasaki}}, \bibnamefont{and}
  \bibinfo{author}{\bibfnamefont{T.}~\bibnamefont{Tanaka}},
  \bibinfo{journal}{Phys. Rev. D} \textbf{\bibinfo{volume}{55}},
  \bibinfo{pages}{3457} (\bibinfo{year}{1997}).

\bibitem[{\citenamefont{Quinn and Wald}(1997)}]{QuinnWald97}
\bibinfo{author}{\bibfnamefont{T.~C.} \bibnamefont{Quinn}} \bibnamefont{and}
  \bibinfo{author}{\bibfnamefont{R.~M.} \bibnamefont{Wald}},
  \bibinfo{journal}{Phys. Rev. D} \textbf{\bibinfo{volume}{56}},
  \bibinfo{pages}{3381} (\bibinfo{year}{1997}).

\bibitem[{\citenamefont{Quinn}(2000)}]{Quinn00}
\bibinfo{author}{\bibfnamefont{T.~C.} \bibnamefont{Quinn}},
  \bibinfo{journal}{Phys. Rev. D} \textbf{\bibinfo{volume}{62}},
  \bibinfo{pages}{064029} (\bibinfo{year}{2000}).

\bibitem[{\citenamefont{Thorne and Hartle}(1985)}]{ThorneHartle85}
\bibinfo{author}{\bibfnamefont{K.~S.} \bibnamefont{Thorne}} \bibnamefont{and}
  \bibinfo{author}{\bibfnamefont{J.~B.} \bibnamefont{Hartle}},
  \bibinfo{journal}{Phys. Rev. D} \textbf{\bibinfo{volume}{31}},
  \bibinfo{pages}{1815} (\bibinfo{year}{1985}).

\bibitem[{\citenamefont{Zhang}(1986)}]{Zhang86}
\bibinfo{author}{\bibfnamefont{X.-H.} \bibnamefont{Zhang}},
  \bibinfo{journal}{Phys. Rev. D} \textbf{\bibinfo{volume}{34}},
  \bibinfo{pages}{991} (\bibinfo{year}{1986}).

\bibitem[{\citenamefont{Synge}(1960)}]{Synge60}
\bibinfo{author}{\bibnamefont{Synge}}, \emph{\bibinfo{title}{Relativity: The
  General Theory}} (\bibinfo{publisher}{North Holland},
  \bibinfo{address}{Amsterdam}, \bibinfo{year}{1960}).

\bibitem[{\citenamefont{Barack and Ori}(2000)}]{BarackOri00}
\bibinfo{author}{\bibfnamefont{L.}~\bibnamefont{Barack}} \bibnamefont{and}
  \bibinfo{author}{\bibfnamefont{A.}~\bibnamefont{Ori}},
  \bibinfo{journal}{Phys. Rev. D} \textbf{\bibinfo{volume}{61}},
  \bibinfo{pages}{061502(R)} (\bibinfo{year}{2000}).

\bibitem[{\citenamefont{Barack and Ori}(2002)}]{BarackOri02}
\bibinfo{author}{\bibfnamefont{L.}~\bibnamefont{Barack}} \bibnamefont{and}
  \bibinfo{author}{\bibfnamefont{A.}~\bibnamefont{Ori}},
  \bibinfo{journal}{Phys. Rev. D} \textbf{\bibinfo{volume}{66}},
  \bibinfo{pages}{084022} (\bibinfo{year}{2002}), \eprint{gr-qc/0204093}.

\bibitem[{\citenamefont{Barack et~al.}(2002)\citenamefont{Barack, Mino, Nakano,
  Ori, and Sasaki}}]{BMNOS02}
\bibinfo{author}{\bibfnamefont{L.}~\bibnamefont{Barack}},
  \bibinfo{author}{\bibfnamefont{Y.}~\bibnamefont{Mino}},
  \bibinfo{author}{\bibfnamefont{H.}~\bibnamefont{Nakano}},
  \bibinfo{author}{\bibfnamefont{A.}~\bibnamefont{Ori}}, \bibnamefont{and}
  \bibinfo{author}{\bibfnamefont{M.}~\bibnamefont{Sasaki}},
  \bibinfo{journal}{Phys. Rev. Lett.} \textbf{\bibinfo{volume}{88}},
  \bibinfo{pages}{091101} (\bibinfo{year}{2002}).

\bibitem[{\citenamefont{Mino et~al.}(2002)\citenamefont{Mino, Nakano, and
  Sasaki}}]{Mino02}
\bibinfo{author}{\bibfnamefont{Y.}~\bibnamefont{Mino}},
  \bibinfo{author}{\bibfnamefont{H.}~\bibnamefont{Nakano}}, \bibnamefont{and}
  \bibinfo{author}{\bibfnamefont{M.}~\bibnamefont{Sasaki}},
  \bibinfo{journal}{Prog. Theor. Phys.} \textbf{\bibinfo{volume}{108}},
  \bibinfo{pages}{1039} (\bibinfo{year}{2002}), \eprint{gr-qc/0111074}.

\bibitem[{\citenamefont{Barack and Ori}(2003)}]{BarackOri03}
\bibinfo{author}{\bibfnamefont{L.}~\bibnamefont{Barack}} \bibnamefont{and}
  \bibinfo{author}{\bibfnamefont{A.}~\bibnamefont{Ori}},
  \bibinfo{journal}{Phys. Rev. D} \textbf{\bibinfo{volume}{67}},
  \bibinfo{pages}{024029} (\bibinfo{year}{2003}).

\bibitem[{\citenamefont{Burko}(2000)}]{Burko00}
\bibinfo{author}{\bibfnamefont{L.~M.} \bibnamefont{Burko}},
  \bibinfo{journal}{Phys. Rev. Lett.} \textbf{\bibinfo{volume}{84}},
  \bibinfo{pages}{4529} (\bibinfo{year}{2000}).

\bibitem[{\citenamefont{Barack and Burko}(2000)}]{BarackBurko00}
\bibinfo{author}{\bibfnamefont{L.}~\bibnamefont{Barack}} \bibnamefont{and}
  \bibinfo{author}{\bibfnamefont{L.~M.}~\bibnamefont{Burko}},
  \bibinfo{journal}{Phys. Rev. D} \textbf{\bibinfo{volume}{62}},
  \bibinfo{pages}{084040} (\bibinfo{year}{2000}).

\bibitem[{\citenamefont{Misner et~al.}(1973)\citenamefont{Misner, Thorne, and
  Wheeler}}]{MTW}
\bibinfo{author}{\bibfnamefont{C.~W.} \bibnamefont{Misner}},
  \bibinfo{author}{\bibfnamefont{K.~S.} \bibnamefont{Thorne}},
  \bibnamefont{and} \bibinfo{author}{\bibfnamefont{J.~A.}
  \bibnamefont{Wheeler}}, \emph{\bibinfo{title}{Gravitation}}
  (\bibinfo{publisher}{Freeman}, \bibinfo{address}{San Fransisco},
  \bibinfo{year}{1973}).

\bibitem[{\citenamefont{Thorne and Kov\'{a}cs}(1975)}]{ThorneKovacs75}
\bibinfo{author}{\bibfnamefont{K.~S.} \bibnamefont{Thorne}} \bibnamefont{and}
  \bibinfo{author}{\bibfnamefont{S.~J.} \bibnamefont{Kov\'{a}cs}},
  \bibinfo{journal}{Astrophys. J.} \textbf{\bibinfo{volume}{200}},
  \bibinfo{pages}{245} (\bibinfo{year}{1975}).

\bibitem[{\citenamefont{Damour and Iyer}(1991{\natexlab{a}})}]{DamourIyer91}
\bibinfo{author}{\bibfnamefont{T.}~\bibnamefont{Damour}} \bibnamefont{and}
  \bibinfo{author}{\bibfnamefont{B.~R.} \bibnamefont{Iyer}},
  \bibinfo{journal}{Ann. Inst. H. Poincar\'e (Phys. Theorique)}
  \textbf{\bibinfo{volume}{54}}, \bibinfo{pages}{115}
  (\bibinfo{year}{1991}{\natexlab{a}}).

\bibitem[{\citenamefont{Press et~al.}(1992)\citenamefont{Press, Teukolsky,
  Vetterling, and Flannery}}]{NumericalRecipes}
\bibinfo{author}{\bibfnamefont{W.~H.} \bibnamefont{Press}},
  \bibinfo{author}{\bibfnamefont{S.~A.} \bibnamefont{Teukolsky}},
  \bibinfo{author}{\bibfnamefont{W.~T.} \bibnamefont{Vetterling}},
  \bibnamefont{and} \bibinfo{author}{\bibfnamefont{B.~P.}
  \bibnamefont{Flannery}}, \emph{\bibinfo{title}{Numerical Recipes in {C}: The
  Art of Scientific Computing}} (\bibinfo{publisher}{Cambridge University
  Press}, \bibinfo{address}{Cambridge}, \bibinfo{year}{1992}),
  \bibinfo{edition}{2nd} ed.

\bibitem[{\citenamefont{Detweiler}(2001)}]{Det01}
\bibinfo{author}{\bibfnamefont{S.}~\bibnamefont{Detweiler}},
  \bibinfo{journal}{Phys. Rev. Lett.} \textbf{\bibinfo{volume}{86}},
  \bibinfo{pages}{1931} (\bibinfo{year}{2001}).

\bibitem[{\citenamefont{Lousto}(2000)}]{Lousto00}
\bibinfo{author}{\bibfnamefont{C.~O.} \bibnamefont{Lousto}},
  \bibinfo{journal}{Phys. Rev. Lett.} \textbf{\bibinfo{volume}{84}},
  \bibinfo{pages}{5251} (\bibinfo{year}{2000}).

\bibitem[{\citenamefont{Lousto}(2001)}]{Lousto01}
\bibinfo{author}{\bibfnamefont{C.~O.} \bibnamefont{Lousto}},
  \bibinfo{journal}{Class. Quantum Grav.} \textbf{\bibinfo{volume}{18}},
  \bibinfo{pages}{3989} (\bibinfo{year}{2001}).

\bibitem[{\citenamefont{Barack and Lousto}(2002)}]{BarackLousto02}
\bibinfo{author}{\bibfnamefont{L.}~\bibnamefont{Barack}} \bibnamefont{and}
  \bibinfo{author}{\bibfnamefont{C.~O.} \bibnamefont{Lousto}},
  \bibinfo{journal}{Phys. Rev. D} \textbf{\bibinfo{volume}{66}},
  \bibinfo{pages}{061502} (\bibinfo{year}{2002}), \eprint{gr-qc/0205043}.

\bibitem[{\citenamefont{Weinberg}(1972)}]{Weinberg}
\bibinfo{author}{\bibfnamefont{S.}~\bibnamefont{Weinberg}},
  \emph{\bibinfo{title}{Gravitation and Cosmology}}
  (\bibinfo{publisher}{Wiley}, \bibinfo{address}{New York},
  \bibinfo{year}{1972}).

\bibitem[{\citenamefont{Damour and Iyer}(1991{\natexlab{b}})}]{DamourIyer91b}
\bibinfo{author}{\bibfnamefont{T.}~\bibnamefont{Damour}} \bibnamefont{and}
  \bibinfo{author}{\bibfnamefont{B.~R.} \bibnamefont{Iyer}},
  \bibinfo{journal}{Phys. Rev. D} \textbf{\bibinfo{volume}{43}},
  \bibinfo{pages}{3259} (\bibinfo{year}{1991}{\natexlab{b}}).

\bibitem[{\citenamefont{Abramowitz and Stegun}(1965)}]{AandS}
\bibinfo{editor}{\bibfnamefont{M.}~\bibnamefont{Abramowitz}} \bibnamefont{and}
  \bibinfo{editor}{\bibfnamefont{I.~A.} \bibnamefont{Stegun}}, eds.,
  \emph{\bibinfo{title}{Handbook of Mathematical Functions}}
  (\bibinfo{publisher}{Dover Publications}, \bibinfo{address}{New York},
  \bibinfo{year}{1965}).

\bibitem[{\citenamefont{Arfken and Weber}(2001)}]{Arfken}
\bibinfo{author}{\bibfnamefont{G.~B.} \bibnamefont{Arfken}} \bibnamefont{and}
  \bibinfo{author}{\bibfnamefont{H.~J.} \bibnamefont{Weber}},
  \emph{\bibinfo{title}{Mathematical Methods for Physicists}}
  (\bibinfo{publisher}{Harcourt/Academic Press}, \bibinfo{address}{San Diego},
  \bibinfo{year}{2001}), \bibinfo{edition}{5th} ed.

\bibitem[{\citenamefont{Breuer et~al.}(1973)\citenamefont{Breuer, Chrzanowski,
  {Hughes III}, and Misner}}]{Breuer73}
\bibinfo{author}{\bibfnamefont{R.~A.} \bibnamefont{Breuer}},
  \bibinfo{author}{\bibfnamefont{P.~L.} \bibnamefont{Chrzanowski}},
  \bibinfo{author}{\bibfnamefont{H.~G.} \bibnamefont{{Hughes III}}},
  \bibnamefont{and} \bibinfo{author}{\bibfnamefont{C.~W.}
  \bibnamefont{Misner}}, \bibinfo{journal}{Phys. Rev. D}
  \textbf{\bibinfo{volume}{8}}, \bibinfo{pages}{4309} (\bibinfo{year}{1973}).

\end{thebibliography}

\end{document}